# FRACTIONAL OPTIMAL CONTROL MODEL OF SARS-CoV-2 (COVID-19) DISEASE IN GHANA

Samuel Okyere[1*], Joseph Ackora-Prah[1], Kwaku Forkuoh Darkwah[1], Francis Tabi Oduro[2], and Ebenezer Bonyah[3,4]

[1]Department of Mathematics, Kwame Nkrumah University of Science and Technology, Kumasi, Ghana

[2]African Institute of Mathematical Sciences, Ghana
[3]Department of Mathematics Education, Akenten Appiah-Menka University of Skills Training and Enterpreneurial Development, Kumasi, Ghana

[4]Department of Mathematics, Faculty of Science and Technology, Universitas Airlangga, Surabaya 60115, Indonesia.

Corresponding Author: Samuel Okyere; okyere2015@gmail.com

**Abstract**
Research focus on optimal control problems brought on by fractional differential equations has been extensively applied in practice. However, because they are still open-ended and challenging, a number of problems with fractional mathematical modeling and problems with optimal control require additional study. Using fractional-order derivatives defined in the Atangana-Baleanu-Caputo sense, we alter the integer-order model that has been proposed in the literature. We prove the solution's existence, uniqueness, equilibrium points, fundamental reproduction number, and local stability of the equilibrium points. The operator's numerical approach was put into practice to obtain a numerical simulation to back up the analytical conclusions. Fractional optimum controls were incorporated into the model to identify the most efficient intervention strategies for controlling the disease. Utilizing actual data from Ghana for the months of March 2020 to March 2021, the model is validated. The simulation's results show that the fractional operator significantly affected each compartment and that the incidence rate of the population rose when $\nu \geq 0.6$. The examination of the most effective control technique discovered that social exclusion and vaccination were both very effective methods for halting the development of the illness.

## 1 Introduction

The large family of viruses known as coronaviruses is responsible for a number of illnesses, including the Middle East Respiratory Syndrome (MERS), common cold, and severe acute respiratory syndrome (SARS-CoV-2). Never previously have humans been exposed to this novel coronavirus strain [1]. When an infected person coughs, breathes, sneezes, or talks, tiny droplets or particles, such as aerosols, are discharged into the air and transmit the disease [2, 3]. The illness's usual symptoms include fever, coughing, exhaustion, shortness of breath or other breathing problems, as well as loss of taste and smell [4-6]. Over 6 million people had died and there were over 513 million confirmed infections worldwide as of May 1, 2022. Africa as a whole has recorded 11 million confirmed cases, with Ghana accounting for 161,173 of those cases [7].

Numerous scholars have created models for the realization and control of the spread of transmissible illnesses in a population [8]. The representation of infectious diseases using fractional calculus has attracted a lot of attention lately. Examples include malaria [9], TB [10, 11], syphilis [12], chickenpox [13], and most recently COVID-19 [14–24]. Fractional calculus, which is a generalization of differentiation and integration of integer orders, has been proposed to address several of the constraints of integer order derivatives [14]. The fractional order may be able to depict more complex dynamics and include memory effects, which are prevalent in many real-world occurrences, in comparison to the integer model [10, 15]. Using data from China, Italy, and France, Bahloul et al. [16] proposed a fractional-order SEIQRDP model to study the COVID-19 pandemic. To examine the illness transmission in Spain, researchers in Ref. [17] proposed a new SEIRS dynamical model that uses the fractional-order derivative and adds the vaccine rate. The authors of Ref. [18] presented a Caputo fractional SIR epidemic model taking a nonlinear incidence rate into consideration. Regarding the fractal-fractional Atangana-Baleanu derivative, Khan and Atangana [19] considered a fractional model to describe the transmission of COVID-19 while accounting for isolation and quarantine of individuals.

There are several real-world uses for research on optimum control problems brought on by fractional differential equations. However, there are a number of difficulties and problems with fractional mathematical modeling as well as challenges with optimal control that need further study. A general formulation for the optimal control problem for a family of fuzzy fractional differential systems connected to SIR and SEIR epidemic models was developed by Dong et al. [25] using real data from Italy and South Korea. Khan et al. [21] explored a fractional COVID-19 epidemic model that included fractional optimum control and had a convex incidence rate in the sense of Atangana-Baleanu and Caputo. Baba and Bilgehan [22] devised a fractional optimum control issue that incorporates public awareness and treatment for the disease outbreak using a mathematical model with fractional order derivative in the Caputo sense. Nabi et al. [23] created a compartmental model combining all workable non-pharmaceutical intervention options using the classical and Caputo-Fabrizio fractional-order derivatives to study illness transmission in Bangladesh and India. The age structure and fractional-order derivatives were used to create a more accurate version of the conventional SEIR model [24]. They expanded on their methods by include follow-up controls, diagnostics, and awareness programs. In Japan, Das and Samanta [25] suggested a Susceptible - Asymptomatic - Infectious - Recovered (SAIR) compartmental model inside a fractional-order framework that included the best possible management of social distance.

We propose a fractional-order derivative defined in the Atangana - Baleanu - Caputo sense in the current study to investigate the model presented in [20]. The nonlocal characteristic of the virus dynamics is not sufficiently captured by the classical model proposed in [20]. Because the Atangana - Baleanu and Caputo derivatives have a number of desired properties, such as nonlocality and nonsingularity in their kernels, and because this operator can only accurately represent the crossover behavior of the model, it was decided to utilize them to design the model. Other operators without similar qualities, such as Caputo and Caputo-Fabrizio, may or may not adequately explain the dynamics of the coronavirus [19]. Several articles using the Atangana-Baleanu-Caputo derivation are linked and can be found in [13, 26–28].

The following sections then make up the remainder of the paper: In Section 2, we create and analyze a mathematical model that makes use of the fractional-order derivative as established by Atangana, Baleanu, and Caputo. Section 3 identifies the qualitative traits of the model. We identify equilibrium locations, their stability, and the fundamental reproduction number. We incorporate time-dependent optimal control into the constructed model and analyze the optimal control model in Section 4. The numerical framework of the fractional-order model is provided in Section 5, and the numerical analysis is then presented in Section 6. In section 7, the numerical investigation of the optimal control model. Finally, in Section 8, we explain and illustrate the results of our suggested model.

## 2 Model Formulation

In this section, utilizing the fractional-order derivatives derived in the Atangana-Baleanu in Caputo sense, we alter the model given in [20] to incorporate a compartment for quarantine individuals. The fractional order operator is defined as $\nu$, where $0 < \nu \leq 1$. Susceptible individuals (S), Exposed (E), Asymptomatic $(I_A)$, Symptomatic $(I_S)$, Vaccinated (V), Quarantined (Q), and Recovered (R) are the seven classifications that make up the model. The main premise of the model formulation is that, in contrast to [20], where only the asymptomatic transmit the virus, both symptomatic and asymptomatic persons spread the virus when they come into touch with susceptible individuals. Other presumptions in [20] are true.

The susceptible population are recruited at the rate $\Omega$ and die at a rate $\mu$. These individuals get exposed to the disease when they come in contact with the asymptomatic and symptomatic at a rate $\beta$. After being exposed to the disease, they either progress to the asymptomatic class at the rate $(1-\alpha)\varphi$ or the symptomatic class at the rate $\alpha\varphi$. Both asymptomatic and symptomatic get quarantined at the rate $\rho$ and $\tau$ respectively. Those vaccinated according to this model, don't get infected but may join the susceptible class at a rate $\Gamma$. The parameter $\mu$ and $\delta$, are the natural and the disease-induced death rate, respectively. The parameters $\sigma, \theta$ and $\gamma$ are the rate of recovery for the asymptomatic, symptomatic, and quarantine class, respectively. The schematic diagram of the model is displayed in Fig. 1.

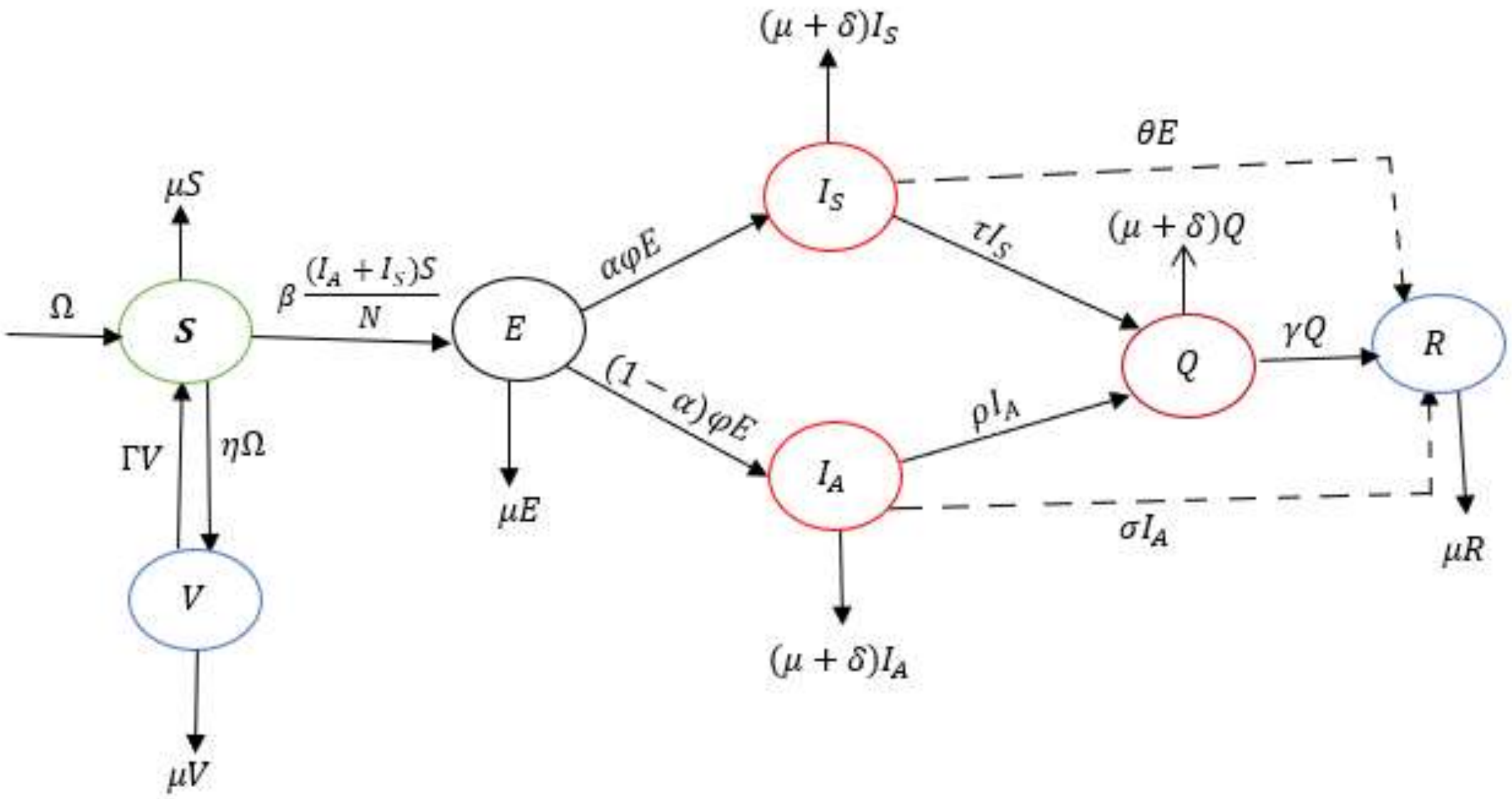

Fig.1: Flowchart of the COVID-19 fractional model

The following fractional derivatives describe the model.

$$
\begin{aligned}
{}_{v}^{ABC}D_t^{v}S &= (1-\eta^{v})\Omega^{v} + \Gamma^{v}V - \beta^{v}S\left(\frac{I_A + I_S}{N}\right) - \mu^{v}S,\\
{}_{v}^{ABC}D_t^{v}E &= \beta^{v}S\left(\frac{I_A + I_S}{N}\right) - (\varphi^{v} + \mu^{v})E,\\
{}_{v}^{ABC}D_t^{v}I_A &= (1-\alpha)\varphi^{v}E - (\delta^{v} + \rho^{v} + \mu^{v} + \sigma^{v})I_A,\\
{}_{v}^{ABC}D_t^{v}I_S &= \alpha\varphi^{v}E - (\theta^{v} + \tau^{v} + \delta^{v} + \mu^{v})I_S,\\
{}_{v}^{ABC}D_t^{v}Q &= \rho^{v}I_A + \tau^{v}I_S - (\gamma^{v} + \mu^{v} + \delta^{v})Q,\\
{}_{v}^{ABC}D_t^{v}V &= \eta^{v}\Omega^{v} - (\Gamma^{v} + \mu^{v})V,\\
{}_{v}^{ABC}D_t^{v}R &= \theta^{v}I_S + \sigma^{v}I_A + \gamma^{v}Q - \mu^{v}R,
\end{aligned}
\tag{1}
$$

with initial conditions $S(0) \geq 0, E(0) \geq 0, I_A(0) \geq 0, I_S(0) \geq 0, Q(0) \geq 0, V(0) \geq 0$ and $R(0) \geq 0$

**2.1 Preliminaries**

We go over the definitions of the key terms used in this work and those specified in [10, 30] in this part.

**Definition 1**

Liouville and Caputo (LC) describe the fractional derivative of order $v$ as in [10, 32] as

$$ {}_{v}^{C}D_{t}^{v}h(t) = \frac{1}{\Gamma(1-v)}\int_{0}^{t}(t-p)^{-v}h(p)dp, \qquad 0 < v \leq 1. \tag{2} $$

**Definition 2**

We provide the Liouville-Caputo sense definition of the Atangana-Baleanu fractional derivative [10, 30]:

$$ {}_{v}^{ABC}D_{t}^{v}h(t) = \frac{B(v)}{(1-v)}\int_{0}^{t}E_{v}\left(-v\left(\frac{(t-p)^{v}}{1-v}\right)\right)\dot{h}(p)dp, \tag{3} $$

where $B(v) = 1 - v + \frac{v}{\Gamma(v)}$, is the normalized function.

**Definition 3**

The Atangana-Baleanu-Caputo derivative's pertinent fractional integral is given by the definition at [10, 30]

$$ {}_{v}^{ABC}I_{t}^{v}h(t) = \frac{(1-v)}{B(v)}h(t) + \frac{v}{B(v)\Gamma(v)}\int_{0}^{t}(t-p)^{v-1}\dot{h}(p)dp. \tag{4} $$

They calculated both derivatives' Laplace transforms and discovered the following:

$$ L\left\{{}_{0}^{ABC}D_{0,t}^{v}h(t)\right\} = \frac{B(v)H(q)q^{v} - q^{v-1}h(0)}{(1-v)\left(q^{v} + \frac{v}{1-v}\right)}. \tag{5} $$

**Theorem 1**

For a function $h \in C[a,b]$, the following results holds [10, 33]:

$$ \left\| {}_{v}^{ABC}D_{t}^{v}r(t) \right\| < \frac{B(v)}{(1-v)}\left\|h(t)\right\|, \text{ where } \left\|h(t)\right\| = \max_{a\leq t\leq b}\left|h(t)\right|. \tag{6} $$

Additionally, the derivatives of Atangana, Baleanu, and Caputo satisfy the Lipschitz criterion [10, 33]:

$$\left\|{}_{v}^{ABC}D_{,t}^{v}h_1(t) - {}_{v}^{ABC}D_{t}^{v}h_2(t)\right\| < \omega\left\|h_1(t) - h_2(t)\right\|. \qquad (7)$$

## 2.2 Existence and Uniqueness

This section establishes the existence and distinctiveness of the solutions to the system (1).

We denote a Banach space by $D(G)$ with $G = [0, b]$, containing real valued continuous function with sup norm $W = D(G)\times D(G)\times D(G)\times D(G)\times D(G)\times D(G)\times D(G)$ and the given norm $\|(S, E, I_A, I_S, Q, V, R)\| = \|S\| + \|E\| + \|I_A\| + \|I_S\| + \|Q\| + \|V\| + \|R\|$, where $\|S\| = Sup_{t\in G}|S|, \|E\| = Sup_{t\in G}|E|, \|I_A\| = Sup_{t\in G}|I_A|, \|I_S\| = Sup_{t\in G}|I_S|, \|Q\| = Sup_{t\in G}|Q|, \|V\| = Sup_{t\in G}|V|, \|R\| = Sup_{t\in G}|R|$. Using the $ABC$ integral operator on the system (1), we have

$$\begin{cases} S(t)-S(0) = {}_{v}^{ABC}D_t^v S(t)\left\{(1-\eta^v)\Omega^v + \Gamma^v V - \beta^v\left(\dfrac{I_A(t)+I_S(t)}{N(t)}\right)S(t) - \mu^v S(t)\right\}, \\ E(t)-E(0) = {}_{v}^{ABC}D_t^v E(t)\left\{\beta^v\left(\dfrac{I_A(t)+I_S(t)}{N(t)}\right)S(t) - (\varphi^v+\mu^v)E(t)\right\}, \\ I_A(t)-I_A(0) = {}_{v}^{ABC}D_t^v I_A(t)\{(1-\alpha)\varphi^v E - (\delta^v+\rho^v+\mu^v+\sigma^v)I_A(t)\}, \\ I_S(t)-I_S(0) = {}_{v}^{ABC}D_t^v I_S(t)\{\alpha\varphi^v E(t) - (\delta^v+\mu^v+\theta^v+\tau^v)I_S(t)\}, \\ Q(t)-Q(0) = {}_{v}^{ABC}D_t^v Q(t)\{\rho^v I_A + \tau^v I_S - (\delta^v+\mu^v+\gamma^v)Q(t)\}, \\ V(t)-V(0) = {}_{v}^{ABC}D_t^v V(t)\{\eta^v\Omega - (\Gamma^v+\mu^v)V(t)\}, \\ R(t)-R(0) = {}_{v}^{ABC}D_t^v R(t)\{\gamma^v Q(t) + \theta^v I_S(t) + \sigma^v I_A(t) - \mu^v R(t)\}. \end{cases} \qquad (8)$$

Now from Definition 1, we have

$$\begin{cases} S(t)-S(0) = \dfrac{1-v}{B(v)}\Phi_1(v,t,S(t)) + \dfrac{v}{B(v)\Gamma(v)}\times\displaystyle\int_0^t (t-\tau)^{v-1}\Phi_1(v,\tau,S(\tau))d\tau, \\ E(t)-E(0) = \dfrac{1-v}{B(v)}\Phi_2(v,t,E(t)) + \dfrac{v}{B(v)\Gamma(v)}\times\displaystyle\int_0^t (t-\tau)^{v-1}\Phi_2(v,\tau,E(\tau))d\tau, \\ I_A(t)-I_A(0) = \dfrac{1-v}{B(v)}\Phi_3(v,t,I_A(t)) + \dfrac{v}{B(v)\Gamma(v)}\times\displaystyle\int_0^t (t-\tau)^{v-1}\Phi_3(v,\tau,I_A(\tau))d\tau, \\ I_S(t)-I_S(0) = \dfrac{1-v}{B(v)}\Phi_4(v,t,I_S(t)) + \dfrac{v}{B(v)\Gamma(v)}\times\displaystyle\int_0^t (t-\tau)^{v-1}\Phi_4(v,\tau,I_S(\tau))d\tau, \\ Q(t)-Q(0) = \dfrac{1-v}{B(v)}\Phi_5(v,t,Q(t)) + \dfrac{v}{B(v)\Gamma(v)}\times\displaystyle\int_0^t (t-\tau)^{v-1}\Phi_5(v,\tau,Q(\tau))d\tau, \\ V(t)-V(0) = \dfrac{1-v}{B(v)}\Phi_6(v,t,V(t)) + \dfrac{v}{B(v)\Gamma(v)}\times\displaystyle\int_0^t (t-\tau)^{v-1}\Phi_6(v,\tau,V(\tau))d\tau, \\ R(t)-R(0) = \dfrac{1-v}{B(v)}\Phi_7(v,t,R(t)) + \dfrac{v}{B(v)\Gamma(v)}\times\displaystyle\int_0^t (t-\tau)^{v-1}\Phi_7(v,\tau,R(\tau))d\tau, \end{cases}$$

(9)

where

$$\begin{cases} \Phi_1(v,\tau,S(t)) = (1-\eta^v)\Omega^v + \Gamma^v V - \beta^v\left(\dfrac{I_A(t)+I_S(t)}{N(t)}\right)S(t) - \mu^v S(t), \\ \Phi_2(v,\tau,E(t)) = \beta^v\left(\dfrac{I_A(t)+I_S(t)}{N(t)}\right)S(t) - (\varphi^v + \mu^v)E(t), \\ \Phi_3(v,\tau,I_A(t)) = (1-\alpha)\varphi^v E - (\delta^v + \rho^v + \mu^v + \sigma^v)I_A(t), \\ \Phi_4(v,\tau,I_S(t)) = \alpha\varphi^v E(t) - (\delta^v + \mu^v + \theta^v + \tau^v)I_S(t), \\ \Phi_5(v,\tau,Q(t)) = \rho^v I_A + \tau^v I_S - (\delta^v + \mu^v + \gamma^v)Q(t), \\ \Phi_6(v,\tau,V(t)) = \eta^v\Omega - (\Gamma^v + \mu^v)V(t), \\ \Phi_7(v,\tau,R(t)) = \gamma^v Q(t) + \theta^v I_S(t) + \sigma^v I_A(t) - \mu^v R(t). \end{cases} \tag{10}$$

Furthermore, the Atangana-Baleanu-Caputo derivatives fulfill the Lipschitz condition [10, 33] only if $S(t), E(t), I_A(t), I_S(t), Q(t), V(t)$ and $R(t)$ possess an upper bound. Suppose $S(t)$ and $S^*(t)$ are couple functions, then

$$\left\|\Phi_1(v,t,S(t)) - \Phi_1(v,t,S^*(t))\right\| = \left\|-\left[\beta^v\left(\frac{I_A(t)+I_S(t)}{N(t)}\right) - \mu^v\right](S(t)-S^*(t))\right\|. \tag{11}$$

Considering

$$d_1 = \left\|-\left(\beta^v\left(\frac{I_A(t)+I_S(t)}{N(t)}\right) - \mu^v\right)\right\|,$$

Equation (11) simplifies to

$$\left\|\Phi_1(v,t,S(t)) - \Phi_1(v,t,S^*(t))\right\| \le d_1\left\|(S(t)-S^*(t))\right\|. \tag{12}$$

Similarly,

$$\begin{aligned}
&\left\|\Phi_2(v,t,E(t))-\Phi_2(v,t,E^*(t))\right\| \le d_2\left\|(E(t)-E^*(t))\right\|,\\
&\left\|\Phi_3(v,t,I_A(t))-\Phi_3(v,t,I_A^*(t))\right\| \le d_3\left\|(I_A(t)-I_A^*(t))\right\|,\\
&\left\|\Phi_4(v,t,I_S(t))-\Phi_4(v,t,I_S^*(t))\right\| \le d_4\left\|(I_S(t)-I_S^*(t))\right\|,\\
&\left\|\Phi_5(v,t,Q(t))-\Phi_5(v,t,Q^*(t))\right\| \le d_5\left\|(Q(t)-Q^*(t))\right\|,\\
&\left\|\Phi_6(v,t,V(t))-\Phi_6(v,t,V^*(t))\right\| \le d_6\left\|(V(t)-V^*(t))\right\|,\\
&\left\|\Phi_7(v,t,R(t))-\Phi_7(v,t,R^*(t))\right\| \le d_7\left\|(R(t)-R^*(t))\right\|,
\end{aligned} \tag{13}$$

where

$$\begin{aligned}
d_2 &= \left\|-(\varphi^v+\mu^v)\right\|,\\
d_3 &= \left\|-(\alpha\varphi^v+\mu^v+\rho^v+\delta^v+\sigma^v)\right\|,\\
d_4 &= \left\|-(\delta^v+\mu^v+\tau^v+\theta^v)\right\|,\\
d_5 &= \left\|-(\delta^v+\mu^v+\gamma^v)\right\|,\\
d_6 &= \left\|-(\Gamma^v+\mu^v)\right\|,\\
d_7 &= \left\|-\mu^v\right\|,
\end{aligned}$$

hence Lipschitz condition holds. Now taking system (9) in a reiterative manner gives

$$\begin{cases}
S_n(t)-S(0)=\dfrac{1-v}{B(v)}\Phi_1(v,t,S_{n-1}(t))+\dfrac{v}{B(v)\Gamma(v)}\times\displaystyle\int_0^t(t-\vartheta)^{v-1}\Phi_1(v,\vartheta,S_{n-1}(\vartheta))d\vartheta,\\
E_n(t)-E(0)=\dfrac{1-v}{B(v)}\Phi_2(v,t,E_{n-1}(t))+\dfrac{v}{B(v)\Gamma(v)}\times\displaystyle\int_0^t(t-\vartheta)^{v-1}\Phi_2(v,\vartheta,E_{n-1}(\vartheta))d\vartheta,\\
I_{A_n}(t)-I_A(0)=\dfrac{1-v}{B(v)}\Phi_3(v,t,I_{A_{n-1}}(t))+\dfrac{v}{B(v)\Gamma(v)}\times\displaystyle\int_0^t(t-\vartheta)^{v-1}\Phi_3(v,\vartheta,I_{A_{n-1}}(\vartheta))d\vartheta,\\
I_{S_n}(t)-I_S(0)=\dfrac{1-v}{B(v)}\Phi_4(v,t,I_{S_{n-1}}(t))+\dfrac{v}{B(v)\Gamma(v)}\times\displaystyle\int_0^t(t-\vartheta)^{v-1}\Phi_4(v,\vartheta,I_{S_{n-1}}(\vartheta))d\vartheta,\\
Q_n(t)-Q(0)=\dfrac{1-v}{B(v)}\Phi_5(v,t,Q_{n-1}(t))+\dfrac{v}{B(v)\Gamma(v)}\times\displaystyle\int_0^t(t-\vartheta)^{v-1}\Phi_5(v,\tau,Q_{n-1}(\vartheta))d\vartheta,\\
V_n(t)-V(0)=\dfrac{1-v}{B(v)}\Phi_6(v,t,V_{n-1}(t))+\dfrac{v}{B(v)\Gamma(v)}\times\displaystyle\int_0^t(t-\vartheta)^{v-1}\Phi_5(v,\tau,V_{n-1}(\vartheta))d\vartheta,\\
R_n(t)-R(0)=\dfrac{1-v}{B(v)}\Phi_7(v,t,R_{n-1}(t))+\dfrac{v}{B(v)\Gamma(v)}\times\displaystyle\int_0^t(t-\vartheta)^{v-1}\Phi_7(v,\vartheta,R_{n-1}(\vartheta))d\vartheta,
\end{cases} \tag{14}$$

Difference of consecutive terms yields

$$\begin{cases}\Xi_{S_n}(t)=S_n(t)-S_{n-1}(t)=\dfrac{1-v}{B(v)}(\Phi_1(v,t,S_{n-1}(t))-\Phi_1(v,t,S_{n-2}(t)))\\ +\dfrac{v}{B(v)\Gamma(v)}\displaystyle\int_0^t(t-\vartheta)^{v-1}(\Phi_1(v,\vartheta,S_{n-1}(\vartheta))-\Phi_1(v,\tau,S_{n-2}(\vartheta)))d\vartheta,\\ \Xi_{E_n}(t)=E_n(t)-E_{n-1}(t)=\dfrac{1-v}{B(v)}(\Phi_2(v,t,E_{n-1}(t))-\Phi_2(v,t,E_{n-2}(t)))\\ +\dfrac{v}{B(v)\Gamma(v)}\displaystyle\int_0^t(t-\vartheta)^{v-1}(\Phi_2(v,\vartheta,E_{n-1}(\vartheta))-\Phi_2(v,\tau,E_{n-2}(\vartheta)))d\vartheta,\\ \Xi_{I_{An}}(t)=I_{An}(t)-I_{A(n-1)}(t)=\dfrac{1-v}{B(v)}(\Phi_3(v,t,I_{A(n-1)}(t))-\Phi_3(v,t,I_{A(n-2)}(t)))\\ +\dfrac{v}{B(v)\Gamma(v)}\displaystyle\int_0^t(t-\vartheta)^{v-1}(\Phi_3(v,\vartheta,I_{A(n-1)}(\vartheta))-\Phi_3(v,\tau,I_{A(n-2)}(\vartheta)))d\vartheta,\\ \Xi_{I_{Sn}}(t)=I_{Sn}(t)-I_{S(n-1)}(t)=\dfrac{1-v}{B(v)}(\Phi_4(v,t,I_{S(n-1)}(t))-\Phi_4(v,t,I_{S(n-2)}(t)))\\ +\dfrac{v}{B(v)\Gamma(v)}\displaystyle\int_0^t(t-\vartheta)^{v-1}(\Phi_4(v,\vartheta,I_{S(n-1)}(\vartheta))-\Phi_4(v,\tau,I_{S(n-2)}(\vartheta)))d\vartheta,\\ \Xi_{Q_n}(t)=Q_n(t)-Q_{n-1}(t)=\dfrac{1-v}{B(v)}(\Phi_5(v,t,Q_{n-1}(t))-\Phi_5(v,t,Q_{n-2}(t)))\\ +\dfrac{v}{B(v)\Gamma(v)}\displaystyle\int_0^t(t-\vartheta)^{v-1}(\Phi_5(v,\vartheta,Q_{n-1}(\vartheta))-\Phi_5(v,\tau,Q_{n-2}(\vartheta)))d\vartheta,\\ \Xi_{V_n}(t)=V_n(t)-V_{n-1}(t)=\dfrac{1-v}{B(v)}(\Phi_6(v,t,V_{n-1}(t))-\Phi_6(v,t,V_{n-2}(t)))\\ +\dfrac{v}{B(v)\Gamma(v)}\displaystyle\int_0^t(t-\vartheta)^{v-1}(\Phi_6(v,\vartheta,V_{n-1}(\vartheta))-\Phi_6(v,\tau,V_{n-2}(\vartheta)))d\vartheta,\\ \Xi_{R_n}(t)=R_n(t)-R_{n-1}(t)=\dfrac{1-v}{B(v)}(\Phi_7(v,t,R_{n-1}(t))-\Phi_7(v,t,R_{n-2}(t)))\\ +\dfrac{v}{B(v)\Gamma(v)}\displaystyle\int_0^t(t-\vartheta)^{v-1}(\Phi_7(v,\vartheta,R_{n-1}(\vartheta))-\Phi_7(v,\tau,R_{n-2}(\vartheta)))d\vartheta,\end{cases} \quad (15)$$

where $S_n(t)=\sum_{i=0}^{n}\Xi_{S_n}(t), E_n(t)=\sum_{i=0}^{n}\Xi_{E_n}(t), I_{An}(t)=\sum_{i=0}^{n}\Xi_{I_{An}}(t), I_{Sn}(t)=\sum_{i=0}^{n}\Xi_{I_{Sn}}(t),$ $Q_n(t)=\sum_{i=0}^{n}\Xi_{Q_n}(t), V_n(t)=\sum_{i=0}^{n}\Xi_{V_n}(t), R_n(t)=\sum_{i=0}^{n}\Xi_{R_n}(t)$. Taking into consideration equations (12) –

(13) and considering $\Xi_{S_{n-1}}(t) = S_{n-1}(t) - S_{n-2}(t), \Xi_{E_{n-1}}(t) = E_{n-1}(t) - E_{n-2}(t),$
$\Xi_{I_A}(t) = I_{A(n-1)}(t) - I_{A(n-2)}(t), \Pi_{I_S}(t) = I_{S(n-1)}(t) - I_{S(n-2)}(t), \Xi_{V_{n-1}}(t) = V_{n-1}(t) - V_{n-2}(t),$
$\Xi_{R_{n-1}}(t) = R_{n-1}(t) - R_{n-2}(t),$

$$\left\{\begin{array}{l}
\left\|\Xi_{S_n}(t)\right\| \le \frac{1-v}{B(v)} d_1 \left\|\Xi_{S_{n-1}}(t)\right\| \frac{v}{B(v)\Gamma(v)} d_1 \times \int_0^t (t-\vartheta)^{v-1} \left\|\Xi_{S_{n-1}}(\vartheta)\right\| d\vartheta, \\
\left\|\Xi_{E_n}(t)\right\| \le \frac{1-v}{B(v)} d_2 \left\|\Xi_{E_{n-1}}(t)\right\| \frac{v}{B(v)\Gamma(v)} d_2 \times \int_0^t (t-\vartheta)^{v-1} \left\|\Xi_{E_{n-1}}(\vartheta)\right\| d\vartheta, \\
\left\|\Xi_{I_{A_n}}(t)\right\| \le \frac{1-v}{B(v)} d_3 \left\|\Xi_{I_{A_{n-1}}}(t)\right\| \frac{v}{B(v)\Gamma(v)} d_3 \times \int_0^t (t-\vartheta)^{v-1} \left\|\Xi_{I_{A_{n-1}}}(\vartheta)\right\| d\vartheta, \\
\left\|\Xi_{I_{S_n}}(t)\right\| \le \frac{1-v}{B(v)} d_4 \left\|\Xi_{I_{S_{n-1}}}(t)\right\| \frac{v}{B(v)\Gamma(v)} d_4 \times \int_0^t (t-\vartheta)^{v-1} \left\|\Xi_{I_{S_{n-1}}}(\vartheta)\right\| d\vartheta, \\
\left\|\Xi_{Q_n}(t)\right\| \le \frac{1-v}{B(v)} d_5 \left\|\Xi_{Q_{n-1}}(t)\right\| \frac{v}{B(v)\Gamma(v)} d_5 \times \int_0^t (t-\vartheta)^{v-1} \left\|\Xi_{Q_{n-1}}(\vartheta)\right\| d\vartheta, \\
\left\|\Xi_{V_n}(t)\right\| \le \frac{1-v}{B(v)} d_6 \left\|\Xi_{V_{n-1}}(t)\right\| \frac{v}{B(v)\Gamma(v)} d_6 \times \int_0^t (t-\vartheta)^{v-1} \left\|\Xi_{V_{n-1}}(\vartheta)\right\| d\vartheta, \\
\left\|\Xi_{R_n}(t)\right\| \le \frac{1-v}{B(v)} d_7 \left\|\Xi_{R_{n-1}}(t)\right\| \frac{v}{B(v)\Gamma(v)} d_7 \times \int_0^t (t-\vartheta)^{v-1} \left\|\Xi_{R_{n-1}}(\vartheta)\right\| d\vartheta.
\end{array}\right. \tag{16}$$

**Theorem 2:** The system (1) has a unique solution for $t \in [0, b]$ subject to the condition $\frac{1-v}{B(v)} d_i + \frac{v}{B(v)\Gamma(v)} b^v n_i < 1, i = 1,2,3,.......,7$ hold [28].

**Proof:**

Since $S(t), E(t), I_A(t), I_S(t), Q(t), V(t)$ and $R(t)$ are bounded functions and Equation $(12)-(13)$ holds. In a recurring manner (16) reaches

$$\begin{cases} \left\|\Xi_{S_n}(t)\right\| \le \left\|S_o(t)\right\| \left(\dfrac{1-v}{B(v)}d_1 + \dfrac{vb^v}{B(v)\Gamma(v)}d_1\right)^n, \\ \left\|\Xi_{E_n}(t)\right\| \le \left\|E_o(t)\right\| \left(\dfrac{1-v}{B(v)}d_2 + \dfrac{vb^v}{B(v)\Gamma(v)}d_2\right)^n, \\ \left\|\Xi_{I_{nn}}(t)\right\| \le \left\|I_{Ao}(t)\right\| \left(\dfrac{1-v}{B(v)}d_3 + \dfrac{vb^v}{B(v)\Gamma(v)}d_3\right)^n, \\ \left\|\Xi_{I_{S_n}}(t)\right\| \le \left\|I_{So}(t)\right\| \left(\dfrac{1-v}{B(v)}d_4 + \dfrac{vb^v}{B(v)\Gamma(v)}d_4\right)^n, \\ \left\|\Xi_{Q_n}(t)\right\| \le \left\|Q_o(t)\right\| \left(\dfrac{1-v}{B(v)}d_5 + \dfrac{vb^v}{B(v)\Gamma(v)}d_5\right)^n, \\ \left\|\Xi_{V_n}(t)\right\| \le \left\|V_o(t)\right\| \left(\dfrac{1-v}{B(v)}d_6 + \dfrac{vb^v}{B(v)\Gamma(v)}d_6\right)^n, \\ \left\|\Xi_{R_n}(t)\right\| \le \left\|R_o(t)\right\| \left(\dfrac{1-v}{B(v)}d_7 + \dfrac{vb^v}{B(v)\Gamma(v)}d_7\right)^n, \end{cases} \tag{17}$$

and

$\left\|\Xi_{S_n}(t)\right\| \to 0, \left\|\Xi_{E_n}(t)\right\| \to 0, \left\|\Xi_{I_{A_n}}(t)\right\| \to 0, \left\|\Xi_{I_{S_n}}(t)\right\| \to 0, \left\|\Xi_{Q_n}(t)\right\| \to 0, \left\|\Xi_{V_n}(t)\right\| \to 0, \left\|\Xi_{R_n}(t)\right\| \to 0$ as $n \to \infty$. Incorporating the triangular inequality and for any $j$, system (17) yields

$$\begin{cases} \left\|S_{n+j}(t) - S_n(t)\right\| \le \sum_{i=n+1}^{n+j} T_1^{\,j} = \dfrac{T_1^{\,n+1} - T_1^{\,n+k+1}}{1-T_1}, \\ \left\|E_{n+j}(t) - E_n(t)\right\| \le \sum_{i=n+1}^{n+j} T_2^{\,j} = \dfrac{T_2^{\,n+1} - T_2^{\,n+k+1}}{1-T_2}, \\ \left\|I_{A(n+j)}(t) - I_{An}(t)\right\| \le \sum_{i=n+1}^{n+j} T_3^{\,j} = \dfrac{T_3^{\,n+1} - T_3^{\,n+k+1}}{1-T_3}, \\ \left\|I_{S(n+j)}(t) - I_{S_n}(t)\right\| \le \sum_{i=n+1}^{n+j} T_4^{\,j} = \dfrac{T_4^{\,n+1} - T_4^{\,n+k+1}}{1-T_4}, \\ \left\|Q_{n+j}(t) - Q_n(t)\right\| \le \sum_{i=n+1}^{n+j} T_5^{\,j} = \dfrac{T_5^{\,n+1} - T_5^{\,n+k+1}}{1-T_5}, \\ \left\|V_{n+j}(t) - V_n(t)\right\| \le \sum_{i=n+1}^{n+j} T_6^{\,j} = \dfrac{T_6^{\,n+1} - T_6^{\,n+k+1}}{1-T_6}, \\ \left\|R_{n+j}(t) - R_n(t)\right\| \le \sum_{i=n+1}^{n+j} T_7^{\,j} = \dfrac{T_7^{\,n+1} - T_7^{\,n+k+1}}{1-T_7}. \end{cases} \tag{18}$$

where $T_i = \frac{1-v}{B(v)} d_i + \frac{v}{B(v)\Gamma(v)} b^v d_i < 1$. Hence there exists a unique solution for system (1)

## 3 Model Analyses

The disease – free equilibrium $(E_0)$ is the steady state solution where there is no infection in the population. This is given as

$$E_0 = (S^0, E^0, I_A^0, I_S^0, Q^0, V^0, R^0) = \left( \frac{\Omega^v(\Gamma^v + \mu^v(1-\eta^v))}{\mu^v(\mu^v + \Gamma^v)}, 0,0,0,0, \frac{\eta^v \Omega^v}{(\Gamma^v + \mu^v)}, 0 \right). \tag{19}$$

The endemic equilibrium $(E_1)$ of system (1) is represented by $E_1 = (S^*, E^*, V^*, I_A^*, I_S^*, Q^*, R^*)$,

where

$$\begin{cases} S^* = \frac{(1-\eta^v)\Omega^v + \Gamma^v V^*}{\beta^v (I_A^* + I_S^*) + \mu^v}, E^* = \frac{\beta^v S^* (I_A^* + I_S^*)}{\varphi^v + \mu^v}, I_A^* = \frac{(1-\alpha)\varphi^v E^*}{\rho^v + \sigma^v + \mu^v + \delta^v}, \\ I_S^* = \frac{\alpha \varphi^v E^*}{\rho^v + \sigma^v + \mu^v + \delta^v}, Q^* = \frac{\rho^v I_A^* + \tau^v I_S^*}{\gamma^v + \mu^v + \delta^v}, V^* = \frac{\eta^v \Omega^v}{\Gamma^v + \mu^v}, R^* = \frac{\theta^v I_S^* + \sigma^v I_A^* + \gamma^v Q^*}{\mu^v} \end{cases} \tag{20}$$

We now ascertain the basic reproduction number $(R_0)$ of the system (1). The total number of secondary instances that one sick person can bring about over the life of the infection in a society that is entirely susceptible is the basic reproduction number [34]. Using the next generation operator technique, denote F and V as matrices representing the newly produced diseases and the transition terms we discover, respectively

$$\begin{cases} F = \begin{bmatrix} 0 & \beta^v S^0 & \beta^v S^0 & 0 \\ (1-\alpha)\varphi^v & 0 & 0 & 0 \\ \alpha\varphi^v & 0 & 0 & 0 \\ 0 & \rho^v & \tau^v & 0 \end{bmatrix}, \\ V = \begin{bmatrix} \varphi^v + \mu^v & 0 & 0 & 0 \\ 0 & \rho^v + \sigma^v + \mu^v + \delta^v & 0 & 0 \\ 0 & 0 & \theta^v + \tau^v + \mu^v + \delta^v & 0 \\ 0 & 0 & 0 & \gamma^v + \mu^v + \delta^v \end{bmatrix} \end{cases}$$

Now the basic reproductive number is given as the spectra radius of the matrix $FV^{-1}$.

$$R_1 = \frac{\beta^v \Omega^v (\Gamma^v + \mu^v (1-\eta^v))}{\mu^v (\mu^v + \Gamma^v)(\rho^v + \sigma^v + \mu^v + \delta^v)} \quad \text{and} \quad R_2 = \frac{\beta^v \Omega^v (\Gamma^v + \mu^v (1-\eta^v))}{\mu^v (\mu^v + \Gamma^v)(\theta^v + \tau^v + \mu^v + \delta^v)}, \tag{21}$$

represents the reproduction number for system (1)

The necessary conditions for the local stability of the endemic equilibrium are established in Theorem 2.

**Theorem 2**: The disease-free equilibrium is locally asymptotically stable if $R_o < 1$ and unstable for $R_0 > 1$.

Proof:

The Jacobian matrix of system (1) is given as

$$J = \begin{pmatrix} -\mu^v & 0 & -\beta^v S & -\beta^v S & 0 & \Gamma^v & 0 \\ \beta^v (I_A + I_S) & -(\varphi^v + \mu^v) & \beta^v S & \beta^v S & 0 & 0 & 0 \\ 0 & (1-\alpha)\varphi^v & -(\rho^v + \sigma^v + \mu^v + \delta^v) & 0 & 0 & 0 & 0 \\ 0 & \alpha\varphi^v & 0 & -(\theta^v + \tau^v + \mu^v + \delta^v) & 0 & 0 & 0 \\ 0 & 0 & \rho^v & \tau^v & -(\gamma^v + \mu^v + \delta^v) & 0 & 0 \\ 0 & 0 & 0 & 0 & 0 & -(\Gamma^v + \mu^v) & 0 \\ 0 & 0 & \sigma^v & \theta^v & \gamma^v & 0 & -\mu^v \end{pmatrix} \tag{22}$$

The Jacobian matrix evaluated at the disease-free equilibrium point is given as

$$J_{E^0} = \begin{pmatrix} -\mu^v & 0 & -\beta^v S^0 & -\beta^v S^0 & 0 & \Gamma^v & 0 \\ 0 & -(\varphi^v + \mu^v) & \beta^v S^0 & \beta^v S^0 & 0 & 0 & 0 \\ 0 & (1-\alpha)\varphi^v & -(\rho^v + \sigma^v + \mu^v + \delta^v) & 0 & 0 & 0 & 0 \\ 0 & \alpha\varphi^v & 0 & -(\theta^v + \tau^v + \mu^v + \delta^v) & 0 & 0 & 0 \\ 0 & 0 & \rho^v & \tau^v & -(\gamma^v + \mu^v + \delta^v) & 0 & 0 \\ 0 & 0 & 0 & 0 & 0 & -(\Gamma^v + \mu^v) & 0 \\ 0 & 0 & \sigma^v & \theta^v & \gamma^v & 0 & -\mu^v \end{pmatrix} \tag{23}$$

We need to show that all eigenvalues of system (23) are negative. The first four eigenvalues are $-\mu^v$, $-(\gamma^v+\mu^v+\delta^v)$ , $-(\Gamma^v+\mu^v)$ and $-\mu^v$. The others are obtained from the submatrix in system (24) formed by excluding the, first, fifth, sixth and seventh rows and columns of system (23). Hence we have

$$J_{E^0} = \begin{pmatrix} -(\varphi^v+\mu^v) & \beta^v S^0 & \beta^v S^0 \\ (1-\alpha)\varphi^v & -(\rho^v+\sigma^v+\mu^v+\delta^v) & 0 \\ \alpha\varphi^v & 0 & -(\theta^v+\tau^v+\mu^v+\delta^v) \end{pmatrix} \tag{24}$$

The characteristic equation of system (24) is

$$\lambda^3 + A_1\lambda^2 + A_2\lambda + A_3 = 0, \tag{25}$$

where,

$$\begin{aligned} A_1 &= (\varphi^v+\mu^v)+(\rho^v+\sigma^v+\mu^v+\delta^v)+A, \\ A_2 &= (\varphi^v+\mu^v)[(\rho^v+\sigma^v+\mu^v+\delta^v)+A]+(\rho^v+\sigma^v+\mu^v+\delta^v)[(A-\mu^v\varphi^v(\mu^v+\Gamma^v)R_0], \\ A_3 &= (\rho^v+\sigma^v+\mu^v+\delta^v)[(\varphi^v+\mu^v)A+\alpha^v\varphi^v\beta^v S^0]+(1-\alpha)\varphi^v A\beta^v S^0, \\ A &= (\theta^v+\tau^v+\mu^v+\delta^v). \end{aligned}$$

From Routh – Hurwitz stability criterion if the conditions $A_1>0, A_3>0$ and $A_1A_2-A_3>0$ are satisfied, then all the roots of the characteristic equation have negative real part which means stable equilibrium.

## 4 Fractional Optimal Control Problem

We add two control functions, $u_1$ and $u_2$ into the system (1). Where control $u_1$ and $u_2$ are social distancing and vaccination, respectively. We include the time-dependent controls into system (1) and we have

$$
\begin{cases}
{}_{v}^{ABC}D_t^v S = (1-\eta^v)\Omega^v + \Gamma^v V - (1-u_1)\beta^v \dfrac{S(I_A+I_S)}{N} - \mu^v S - u_2 S, \\
{}_{v}^{ABC}D_t^v E = (1-u_1)\beta^v \dfrac{S(I_A+I_S)}{N} - (\varphi^v + \mu^v)E, \\
{}_{v}^{ABC}D_t^v I_A = (1-\alpha)\varphi^v E - (\rho^v + \sigma^v + \mu^v + \delta^v)I_A, \\
{}_{v}^{ABC}D_t^v I_S = \alpha\varphi^v E - (\theta^v + \tau^v + \mu^v + \delta^v)I_S, \\
{}_{v}^{ABC}D_t^v Q = \rho^v I_A + \tau^v I_S - (\gamma^v + \mu^v + \delta^v)Q, \\
{}_{v}^{ABC}D_t^v V = \eta^v \Omega^v - (\Gamma^v + \mu^v)V + u_2 S, \\
{}_{v}^{ABC}D_{0,t}^v [R(t)] = \theta^v I_S + \sigma^v I_A + \gamma^v Q - \mu^v R.
\end{cases}
\tag{26}
$$

The objective function for fixed time $t_f$ is given as

$$
J(u_1,u_2) = \int_0^{t_f} [G_1 S(t) + G_2 E(t) + G_3 I_A(t) + G_4 I_S(t) + G_5 Q(t) + \frac{1}{2}(T_1 u_1^2 + T_2 u_2^2)]dt, \tag{27}
$$

where $T_1$ and $T_2$ are the measure of relative cost of interventions associated with the controls $u_1$ and $u_2$. We find optimal controls $u_1$ and $u_2$ that minimizes the cost function

$$
J(u_1,u_2) = \int_0^{t_f} \varsigma(S,E,I_A,I_S,Q,V,R)dt, \tag{28}
$$

subject to the constraint

$$
{}_{v}^{ABC}D_t^v S(t) = \varsigma_1, {}_{v}^{ABC}D_t^v E(t) = \varsigma_2, {}_{v}^{ABC}D_t^v I_A(t) = \varsigma_3, {}_{v}^{ABC}D_t^v I_S(t) = \varsigma_4, {}_{v}^{ABC}D_t^v Q(t) = \varsigma_5,
$$
$$
{}_{v}^{ABC}D_t^v V(t) = \varsigma_6, {}_{v}^{ABC}D_t^v R(t) = \varsigma_7,
$$

where $\varsigma_i = \varsigma(S,E,I_A,I_S,Q,V,R)$, $i = 1,2,3,........,7$, $\Phi = (u_1,u_2) \mid u_i$ is a Lebsegue measurable on $[0,1]$ such that $0 \le (u_1,u_2) \le 1$, $\forall t \in [0,t_f]$, where $t_f$ is the final time and with initial conditions $S(0) = S_o, E(0) = E_o, I_A(0) = I_{Ao}, I_S(0) = I_{so}, Q(0) = Q_o, V(0) = V_o, R(0) = R_o$.

To define the fractional optimal control, we consider the following modified cost function [10]:

$$J = \int_0^{t_f} \left[ H_v(S, E, I_A, I_S, Q, V, R, u_j, t) - \sum_{i=1}^{7} \lambda_i \varsigma_i (S, E, I_A, I_S, Q, V, R, u_j, t) \right] dt, \quad (29)$$

where $i = 1, \ldots\ldots\ldots, 7$ and $j = 1,2,3$.

For the fractional optimal control, the Hamiltonian is

$$H_v(S, E, I_A, I_S, Q, V, R, u_j, t) = v(S, E, I_A, I_S, Q, V, R, u_j, t) + \sum_{i=1}^{7} \lambda_i \varsigma_i (S, E, I_A, I_S, Q, V, R, u_j, t)), \quad (30)$$

Where $i = 1, \ldots\ldots\ldots, 7$ and $j = 1,2,3$. The following are essential for the formulation of the fractional optimal control [28, 10].

$$\begin{aligned} &{}_v^{ABC}D_t^v \Lambda_S = \frac{\partial H_v}{\partial S}, {}_v^{ABC}D_t^v \Lambda_E = \frac{\partial H_v}{\partial E}, {}_v^{ABC}D_t^v \Lambda_{I_A} = \frac{\partial H_v}{\partial I_A}, {}_v^{ABC}D_t^v \Lambda_{I_S} = \frac{\partial H_v}{\partial I_S}, {}_v^{ABC}D_t^v \Lambda_Q = \frac{\partial H_v}{\partial Q}, \\ &{}_v^{ABC}D_t^v \Lambda_V = \frac{\partial H_v}{\partial V}, {}_v^{ABC}D_t^v \Lambda_R = \frac{\partial H_v}{\partial R}, \end{aligned} \quad (31)$$

$$\begin{aligned} &0 = \frac{\partial H_v}{\partial u_i}, \\ &{}_v^{ABC}D_t^v S = \frac{\partial H_v}{\partial \Lambda_S}, {}_v^{ABC}D_t^v E = \frac{\partial H_v}{\partial \Lambda_E}, {}_v^{ABC}D_t^v I_A = \frac{\partial H_v}{\partial \Lambda_{I_A}}, {}_v^{ABC}D_t^v I_S = \frac{\partial H_v}{\partial \Lambda_{I_S}}, {}_v^{ABC}D_t^v Q = \frac{\partial H_v}{\partial \Lambda_Q}, \\ &{}_v^{ABC}D_t^v V = \frac{\partial H_v}{\partial \Lambda_V}, {}_v^{ABC}D_t^v R = \frac{\partial H_v}{\partial \Lambda_R}. \end{aligned} \quad (32)$$

Moreover,

$$\Lambda_S(t_f) = \Lambda_E(t_f) = \Lambda_{I_A}(t_f) = \Lambda_{I_S}(t_f) = \Lambda_Q(t_f) = \Lambda_V(t_f) = \Lambda_R(t_f) = 0, \quad (33)$$

are the Lagrange multipliers. Equations (31) and (32) provides the necessary conditions for the fractional optimal control in terms of the Hamiltonian for the optimal control problem defined above. The Hamiltonian, H, defined by

$$H = k_1 S^* + k_2 E^* + k_3 I_A^* + k_4 I_S^* + k_5 Q^* + \frac{1}{2}(T_1 u_1^2 + T_2 u_2^2) \tag{34}$$
$$+ {}_v^{ABC} D_t^v \Lambda_S + {}_v^{ABC} D_t^v \Lambda_E + {}_v^{ABC} D_t^v \Lambda_{I_A} + {}_v^{ABC} D_t^v \Lambda_{I_S} + {}_v^{ABC} D_t^v \Lambda_Q + {}_v^{ABC} D_t^v \Lambda_V + {}_v^{ABC} D_t^v \Lambda_R$$

**Theorem 6:** Given an optimal control $(u_1^*, u_2^*)$ and corresponding solution $S^*, E^*, I_A^*, I_S^*, Q^*, V^*, R^*$ of the system (26) – (27) that minimizes $J(u)$ over U, there exist adjoint variables $\Lambda_S, \Lambda_E, \Lambda_{I_A}, \Lambda_{I_S}, \Lambda_Q, \Lambda_V, \Lambda_R$, satisfying [28]

$$-\frac{d\Lambda_i}{dt} = \frac{\partial H}{\partial i}, \tag{35}$$

where $i = S, E, I_A, I_S, Q, V, R$ with the transversality conditions

$$\Lambda_S(t_f) = \Lambda_E(t_f) = \Lambda_{I_A}(t_f) = \Lambda_{I_S}(t_f) = \Lambda_Q(t_f) = \Lambda_V(t_f) = \Lambda_R(t_f) = 0. \tag{36}$$

***Proof:***

The differential equations characterized by the adjoint variables are obtained by considering the right-hand side differentiation of system (34) determined by the optimal control. The adjoint equations derived are given as

$$
\begin{aligned}
&{}_{v}^{ABC}D_t^v \Lambda_S = \beta^v (I_A - I_S)(1-u_1)[\Lambda_S - \Lambda_E] + (\mu^v + u_2)\Lambda_S + u_2\Lambda_{V,}\\
&{}_{v}^{ABC}D_t^v \Lambda_E = (\varphi^v + \mu^v)\Lambda_E - (1-\alpha)\varphi^v \Lambda_{I_A} - \alpha\varphi^v \Lambda_{I_S},\\
&{}_{v}^{ABC}D_t^v \Lambda_{I_A} = (\rho^v + \sigma^v + \mu^v + \delta^v)\Lambda_{I_A} + (1-u_1)\beta^v S\Lambda_S - \rho^v \Lambda_Q - \sigma^v \Lambda_R,\\
&{}_{v}^{ABC}D_t^v \Lambda_{I_S} = (\theta^v + \tau^v + \mu^v + \delta^v)\Lambda_{I_S} + (1-u_1)\beta^v S\Lambda_S - \tau^v \Lambda_Q - \theta^v \Lambda_R,\\
&{}_{v}^{ABC}D_t^v \Lambda_{Q_S} = (\gamma^v + \mu^v + \delta^v)\Lambda_Q - \gamma^v \Lambda_R,\\
&{}_{v}^{ABC}D_t^v \Lambda_{V_S} = -\Gamma^v \Lambda_S + (\Gamma^v + \mu^v)\Lambda_V,\\
&{}_{v}^{ABC}D_t^v \Lambda_{R_S} = \mu^v \Lambda_R.
\end{aligned}
\tag{37}
$$

By obtaining the solution for $u_1^*$ and $u_2^*$ subject to the constraints, we have

$$
\begin{aligned}
&0 = \frac{\partial H}{\partial u_1} = -T_1 u_1 + \beta^v S(I_A + I_S)[\Lambda_E - \Lambda_S],\\
&0 = \frac{\partial H}{\partial u_2} = -T_2 u_2 + S[\Lambda_S - \Lambda_V].
\end{aligned}
\tag{38}
$$

This gives

$$
\begin{aligned}
u_1^* &= min\left(1, \max\left(0, \frac{\beta^v S(I_A + I_S)[\Lambda_E - \Lambda_S]}{T_1}\right)\right)\\
u_2^* &= min\left(1, \max\left(0, \frac{S[\Lambda_S - \Lambda_V]}{T_2}\right)\right)
\end{aligned}
\tag{39}
$$

## 5 Numerical Scheme of the Fractional Derivative

We apply the scheme in [10] to the system (1). Let us consider the first equation of the system (1)

$$
{}_{v}^{ABC}D_t^v [S(t)] = h(t, S(t)), \qquad S(0) = S_o \tag{40}
$$

Applying the fundamental theorem of fractional calculus to equation (40), we obtain

$$
S(t) - S(0) = \frac{1-v}{B(v)} h(t, S(t)) + \frac{v}{\Gamma(v)B(v)} \int_0^t g(\tau, S(\tau))(t-\tau)^{v-1} d\tau \tag{41}
$$

where $B(v) = 1 - v + \dfrac{v}{\Gamma(v)}$ is a normalised function and at $t_{\varepsilon+1}$ we have,

$$S_{\varepsilon+1} = S_o + \frac{(1-v)\Gamma(v)}{(1-v)\Gamma(v)+v} h(t_\varepsilon, S(t_\varepsilon)) + \frac{v}{\Gamma(v)+v(1-\Gamma(v))} \sum_{\partial=0}^{\varepsilon} \int_{t_\partial}^{t_\varepsilon} h \times (t_{\varepsilon+1} - \vartheta)^{v-1} . \tag{42}$$

Implementing two-step Lagrange's interpolation polynomial on the interval $[t_\varepsilon, t_{\varepsilon+1}]$ [10, 31], we have

$$Y = \frac{h(t_\varepsilon, S_\partial)}{g}(\vartheta - t_{\partial-1}) - \frac{h(t_{\partial-1}, S_{\partial-1})}{g}(\vartheta - t_\partial) \tag{43}$$

Equation (43) is replaced with equation (42) and by performing the steps given in [10, 31], we obtain

$$\begin{aligned} S(t_{\varepsilon+1}) &= S(t_0) + \frac{\Gamma(v)(1-v)}{\Gamma(v)(1-v)+v} h(t_\varepsilon, S(t_\varepsilon)) + \frac{1}{(v+1)\Gamma(v)+v} \sum_{\partial=0}^{n} g^v h(t_\partial, S(t_\partial))(\varepsilon+1-\partial)^v \\ &\times (\varepsilon - \partial + 2 + v) - (\varepsilon - \partial)^v (\varepsilon - \partial + 2 + 2v) - g^v h(t_{\partial-1}, S(t_{\partial-1}))(\varepsilon+1-\partial)^{v+1}(\varepsilon - \partial + 2 + v) \\ &- (\varepsilon - \partial)^v (\varepsilon - \partial + 1 + v) \end{aligned} \tag{44}$$

To obtain high stability, we replace the step size $g$ in equation (44) with $\vartheta(g)$ such that $\vartheta(g) = g + O(g^2), \qquad 0 < \vartheta(g) \le 1$ [10, 34].

The new scheme which is called the nonstandard two–step Lagrange interpolation method (NS2LIM) is given as:

$$\begin{aligned} S(t_{\varepsilon+1}) &= S(t_0) + \frac{\Gamma(v)(1-v)}{\Gamma(v)(1-v)+v} h(t_\varepsilon, S(t_\varepsilon)) + \frac{1}{(v+1)(1-v)\Gamma(v)+v} \sum_{\partial=0}^{\varepsilon} \vartheta(g)^v h(t_\partial, S(t_\partial))(\varepsilon+1-\partial)^v \\ &\times (\varepsilon - \partial + 2 + v) - (\varepsilon - \partial)^v (\varepsilon - \partial + 2 + 2v) - \vartheta(g)^v h(t_{\partial-1}, S(t_{\partial-1}))(\varepsilon+1-\partial)^{v+1}(\varepsilon - \partial + 2 + v) \\ &- (\varepsilon - \partial)^v (\varepsilon - \partial + 1 + v) \end{aligned} \tag{45}$$

Similarly,

$$
\begin{aligned}
&E(t_{\varepsilon+1}) = E(t_0) + \frac{\Gamma(v)(1-v)}{\Gamma(v)(1-v)+v} h(t_\varepsilon, E(t_\varepsilon)) + \frac{1}{(v+1)(1-v)\Gamma(v)+v} \sum_{\partial=0}^{\varepsilon} \vartheta(g)^v h(t_\partial, E(t_\partial))(\varepsilon+1-\partial)^v \\
&\times(\varepsilon-\partial+2+v) - (\varepsilon-\partial)^v(\varepsilon-\partial+2+2v) - \vartheta(g)^v h(t_{\partial-1}, E(t_{\partial-1}))(\varepsilon+1-\partial)^{v+1}(\varepsilon-\partial+2+v) \\
&-(\varepsilon-\partial)^v(\varepsilon-\partial+1+v), \\
&I_A(t_{\varepsilon+1}) = I_A(t_0) + \frac{\Gamma(v)(1-v)}{\Gamma(v)(1-v)+v} h(t_\varepsilon, I_A(t_\varepsilon)) + \frac{1}{(v+1)(1-v)\Gamma(v)+v} \sum_{\partial=0}^{\varepsilon} \vartheta(g)^v h(t_\partial, I_A(t_\partial))(\varepsilon+1-\partial)^v \\
&\times(\varepsilon-\partial+2+v) - (\varepsilon-\partial)^v(\varepsilon-\partial+2+2v) - \vartheta(g)^v h(t_{\partial-1}, I_A(t_{\partial-1}))(\varepsilon+1-\partial)^{v+1}(\varepsilon-\partial+2+v) \\
&-(\varepsilon-\partial)^v(\varepsilon-\partial+1+v), \\
&I_S(t_{\varepsilon+1}) = I_S(t_0) + \frac{\Gamma(v)(1-v)}{\Gamma(v)(1-v)+v} h(t_\varepsilon, I_S(t_\varepsilon)) + \frac{1}{(v+1)(1-v)\Gamma(v)+v} \sum_{\partial=0}^{\varepsilon} \vartheta(g)^v h(t_\partial, I_S(t_\partial))(\varepsilon+1-\partial)^v \\
&\times(\varepsilon-\partial+2+v) - (\varepsilon-\partial)^v(\varepsilon-\partial+2+2v) - \vartheta(g)^v h(t_{\partial-1}, I_S(t_{\partial-1}))(\varepsilon+1-\partial)^{v+1}(\varepsilon-\partial+2+v) \\
&-(\varepsilon-\partial)^v(\varepsilon-\partial+1+v), \\
&Q(t_{\varepsilon+1}) = Q(t_0) + \frac{\Gamma(v)(1-v)}{\Gamma(v)(1-v)+v} h(t_\varepsilon, Q(t_\varepsilon)) + \frac{1}{(v+1)(1-v)\Gamma(v)+v} \sum_{\partial=0}^{\varepsilon} \vartheta(g)^v h(t_\partial, Q(t_\partial))(\varepsilon+1-\partial)^v \\
&\times(\varepsilon-\partial+2+v) - (\varepsilon-\partial)^v(\varepsilon-\partial+2+2v) - \vartheta(g)^v h(t_{\partial-1}, Q(t_{\partial-1}))(\varepsilon+1-\partial)^{v+1}(\varepsilon-\partial+2+v) \\
&-(\varepsilon-\partial)^v(\varepsilon-\partial+1+v), \\
&V(t_{\varepsilon+1}) = V(t_0) + \frac{\Gamma(v)(1-v)}{\Gamma(v)(1-v)+v} h(t_\varepsilon, V(t_\varepsilon)) + \frac{1}{(v+1)(1-v)\Gamma(v)+v} \sum_{\partial=0}^{\varepsilon} \vartheta(g)^v h(t_\partial, V(t_\partial))(\varepsilon+1-\partial)^v \\
&\times(\varepsilon-\partial+2+v) - (\varepsilon-\partial)^v(\varepsilon-\partial+2+2v) - \vartheta(g)^v h(t_{\partial-1}, V(t_{\partial-1}))(\varepsilon+1-\partial)^{v+1}(\varepsilon-\partial+2+v) \\
&-(\varepsilon-\partial)^v(\varepsilon-\partial+1+v), \\
&R(t_{\varepsilon+1}) = R(t_0) + \frac{\Gamma(v)(1-v)}{\Gamma(v)(1-v)+v} h(t_\varepsilon, R(t_\varepsilon)) + \frac{1}{(v+1)(1-v)\Gamma(v)+v} \sum_{\partial=0}^{\varepsilon} \vartheta(g)^v h(t_\partial, R(t_\partial))(\varepsilon+1-\partial)^v \\
&\times(\varepsilon-\partial+2+v) - (\varepsilon-\partial)^v(\varepsilon-\partial+2+2v) - \vartheta(g)^v h(t_{\partial-1}, R(t_{\partial-1}))(\varepsilon+1-\partial)^{v+1}(\varepsilon-\partial+2+v) \\
&-(\varepsilon-\partial)^v(\varepsilon-\partial+1+v)
\end{aligned}
\tag{46}
$$

## 5 Numerical Simulation

In this section, we validate the model using the parameter values given in [20] and use the numerical scheme in [10]. The parameter values are given in Table 1.

Table 1: Parameter values and description

| Parameter | Description | Value | Source |
|---|---|---|---|
| $\Omega$ | Recruitment rate | 29.08 | [20, 36] |
| $\beta$ | Transmission rate | 0.9 | [20, 28] |
| $\varphi$ | The rate at which exposed individuals become infectious | 0.25 | [20, 28, 37] |
| $\mu$ | Natural death rate | $0.4252912\times10^{-4}$ | [20, 28] |
| $\delta$ | Disease-induced death rate | $1.6728\times10^{-5}$ | [20, 35] |
| $\theta$ | Recovery rate of symptomatic individuals | $1/14$ | [38] |
| $\Gamma$ | Rate at which vaccinated individuals loses their immunity | $1.52\times10^{-7}$ | [20] |
| $\sigma$ | Recovery rate of asymptomatic individuals | 1/14 | [20, 28] |
| $\gamma$ | Recovery rate of quarantine individuals | 1/14 | [20] |
| $\tau$ | Rate at which symptomatic individuals move to the quarantine class | 0.01 | [38] |
| $\rho$ | Rate at which asymptomatic individuals move to the quarantine class | $1.026\times10^{-7}$ | Assumed |
| $\eta$ | Rate at which susceptible individuals are vaccinated | 0.01624 | [20] |

Using the initial conditions given in [20]: $S(0) = 30800000, E(0) = 0, I_A(0) = 2, I_S(0) = 0,$ $Q(0) = 0, V(0) = 0, R(0) = 0$, and setting the fractional operator $v \in [0.6,1.0]$ at a step-size of $0.1$, the simulations performed are displayed in Figs. 2 – 8. The figures depict the behaviour of all compartments for the first 400 days since the outbreak.

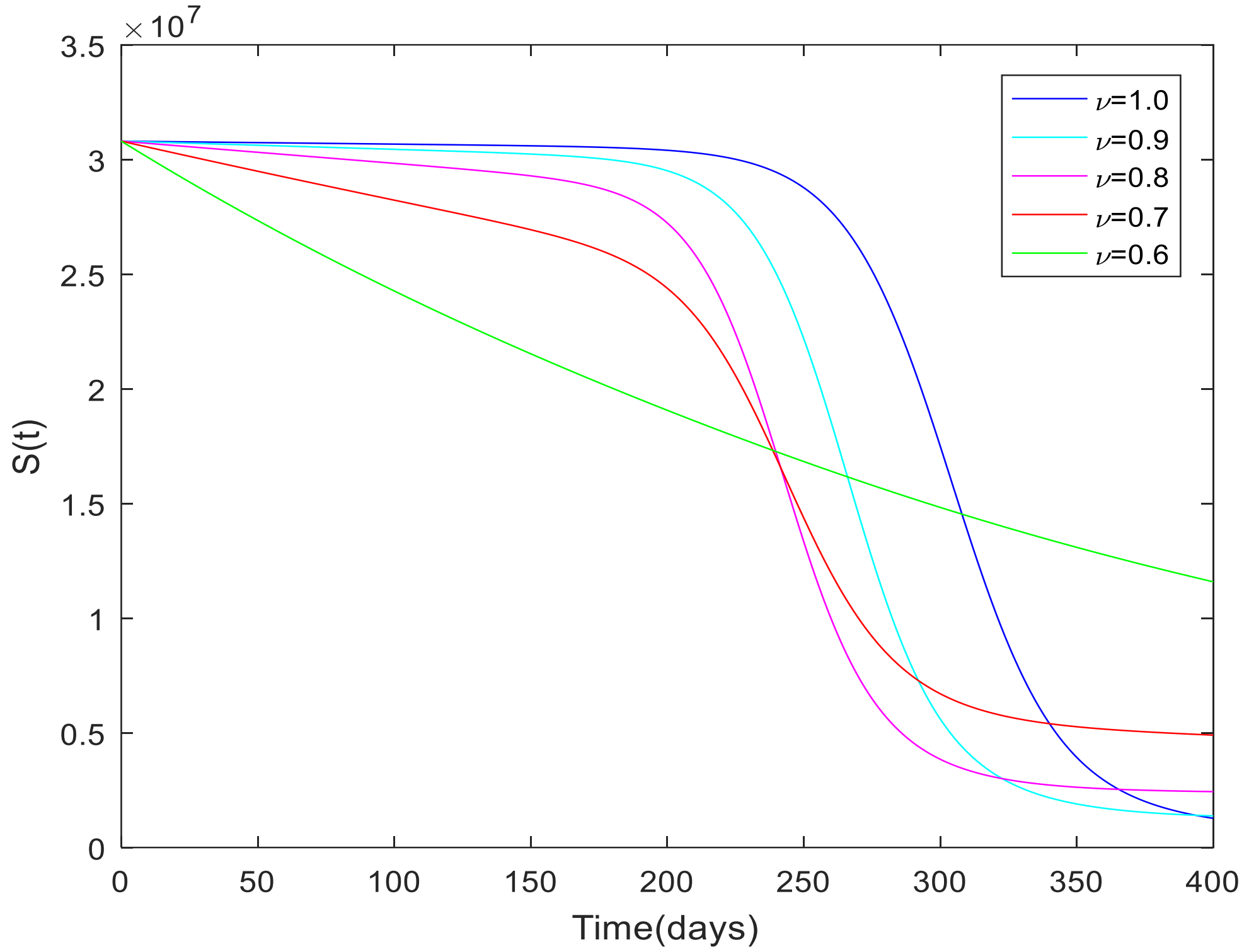

Fig. 2: Behaviour of the susceptible individuals at different values of $v$.

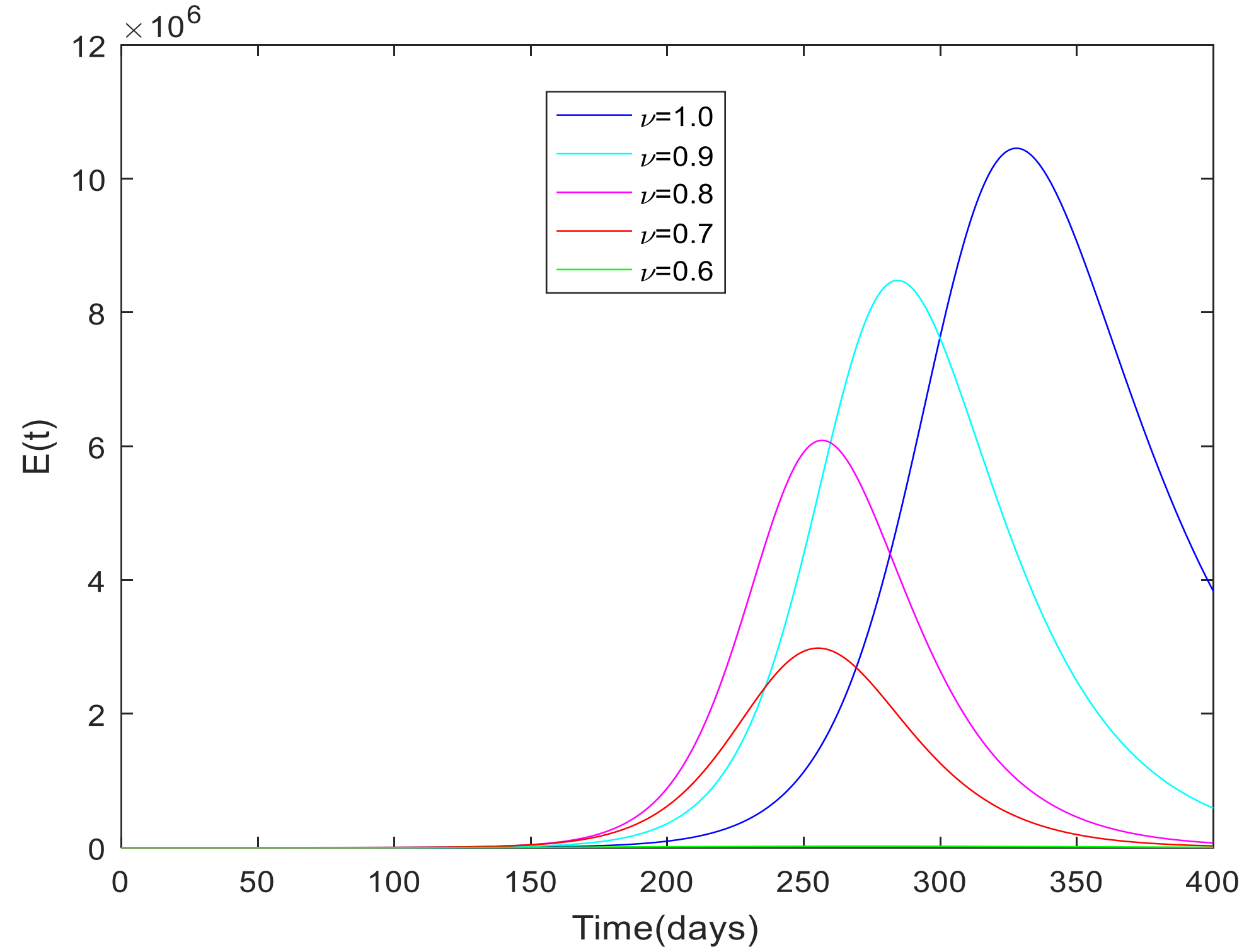

Fig. 3: Behaviour of the exposed individuals at different values of $v$.

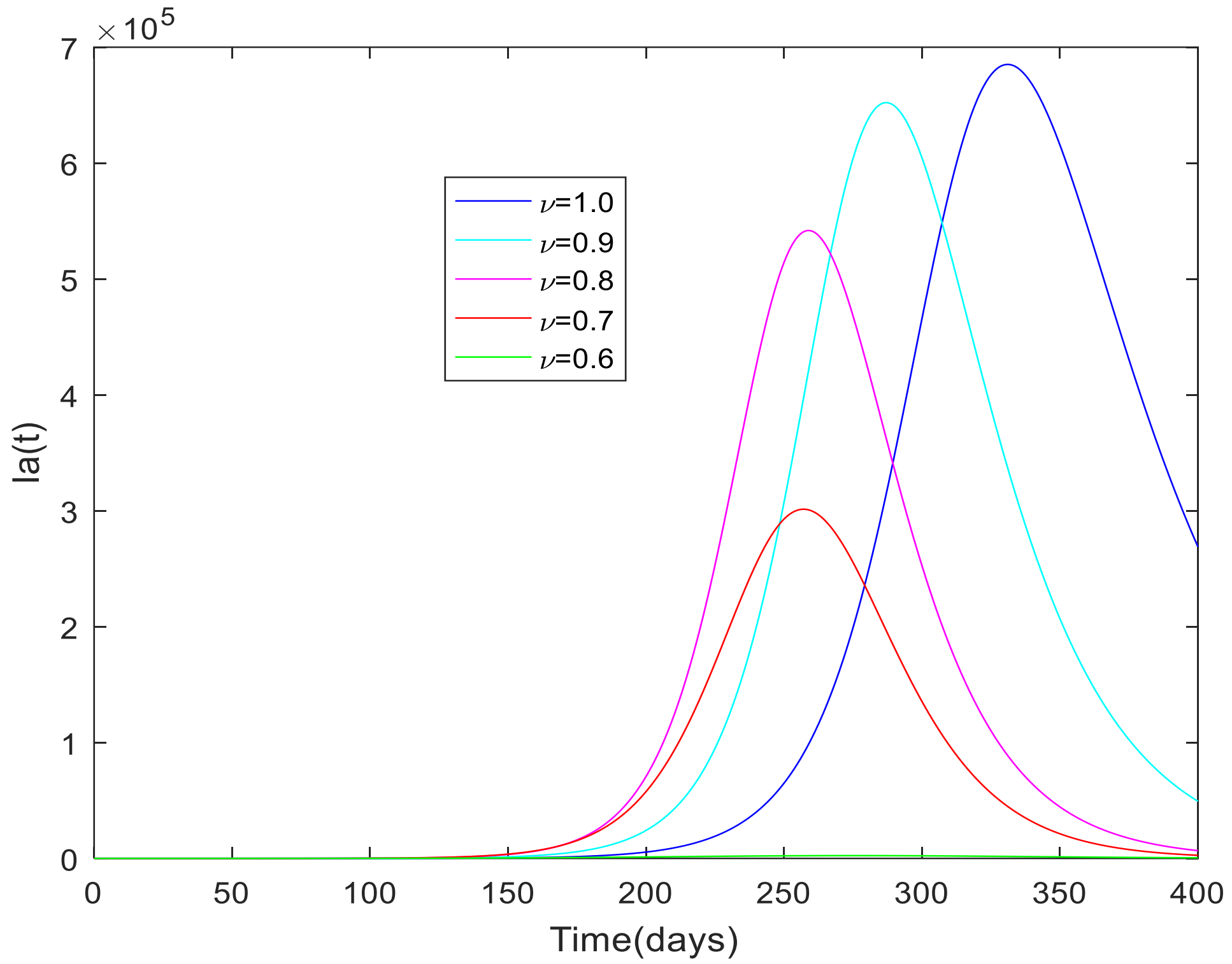

Fig. 4: Behaviour of the asymptomatic at different values of $v$.

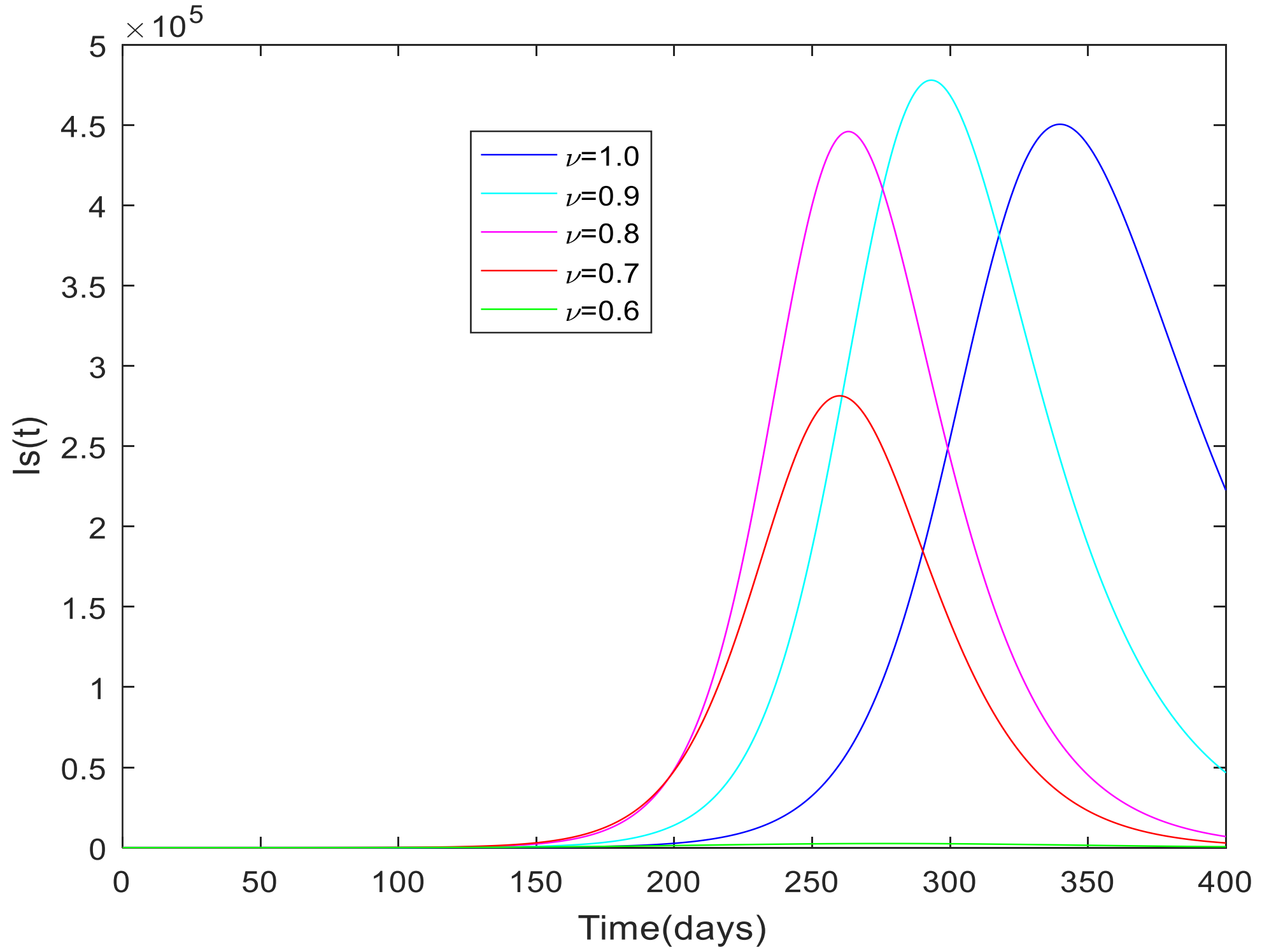

Fig. 5: Behaviour of the symptomatic at different values of $v$.

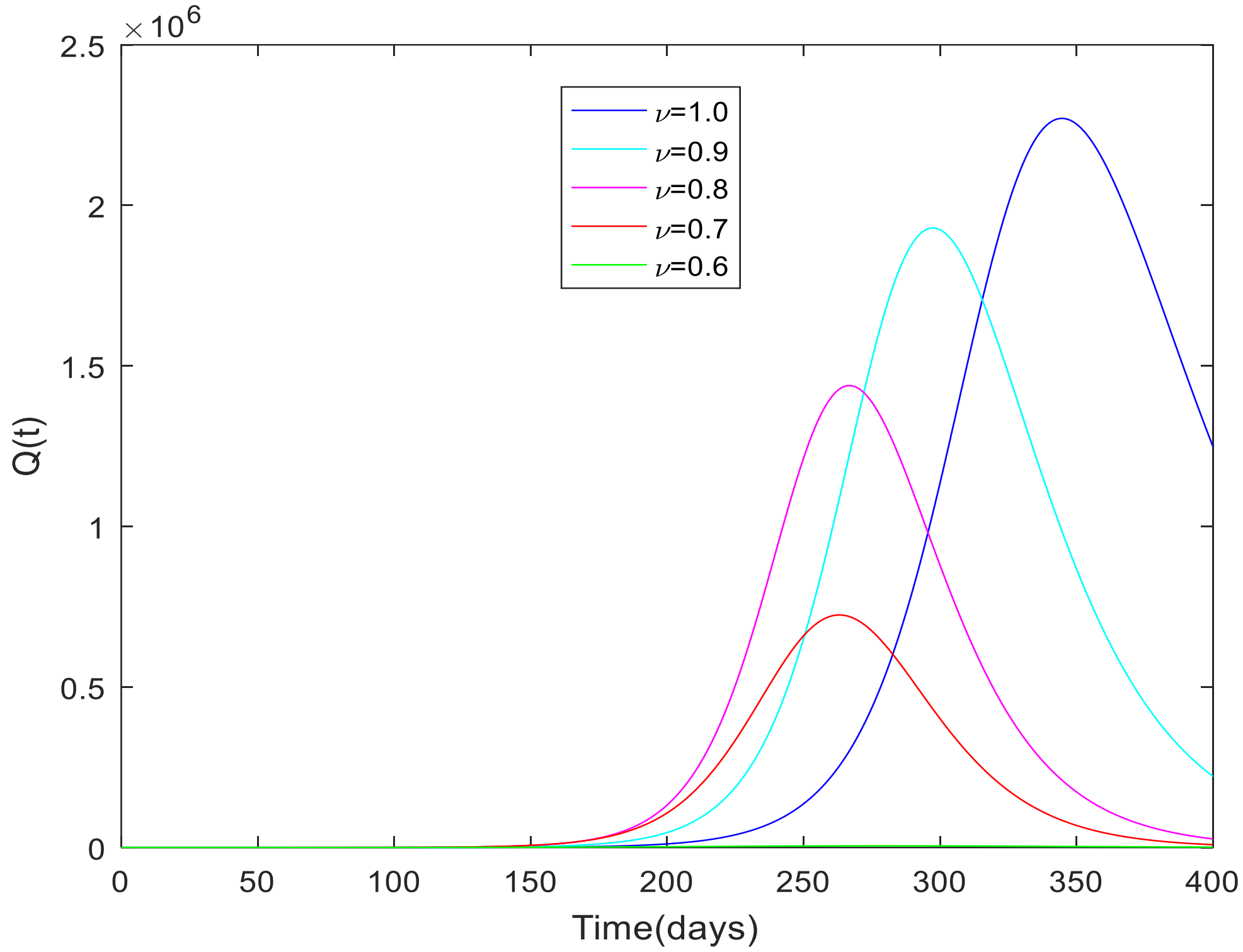

Fig. 6: Behaviour of the quarantine at different values of $\nu$ .

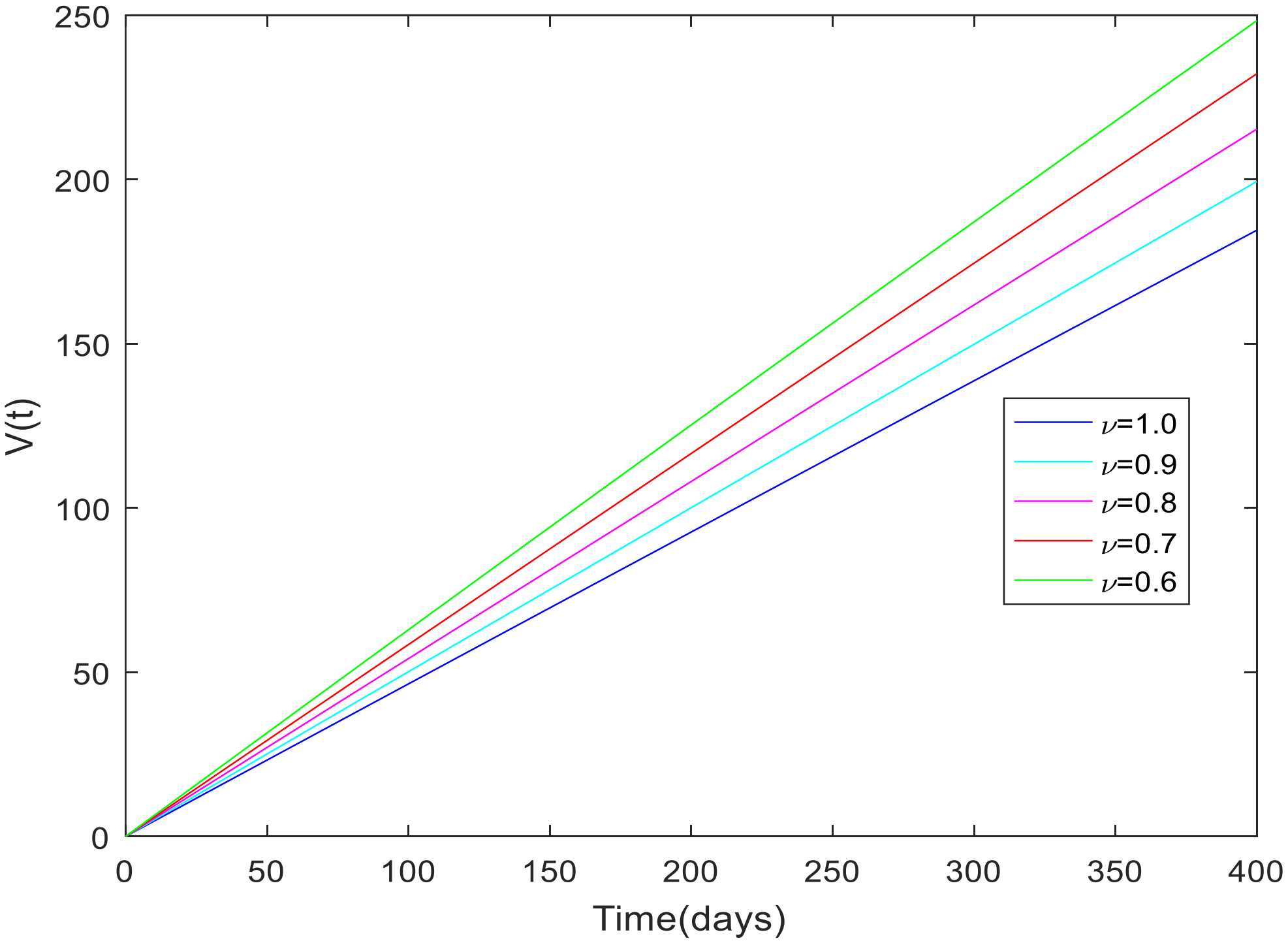

Fig. 7: Behaviour of the vaccinated individuals at different values of $\nu$ .

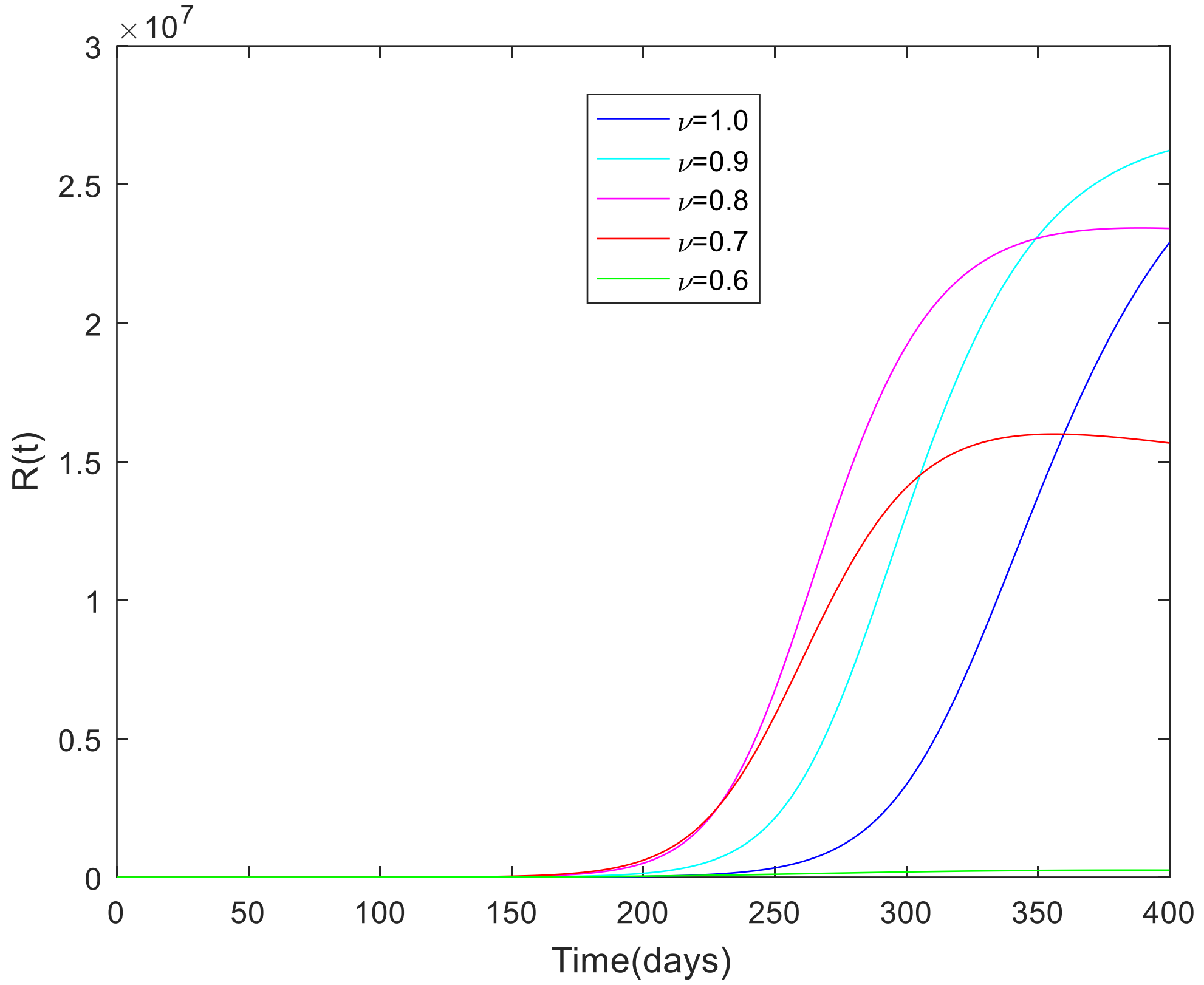

Fig. 8: Behaviour of the recovered individuals at different values of $v$.

Figures 2 – 8 depicts the behaviour of susceptible, exposed, asymptomatic, symptomatic, quarantine, vaccinated, and recovered individuals, respectively, for different values of the fractional operator $v$ for the period of 400 days. In Figure 2, the population of susceptible decreases as the value of the fractional operator $v$, reduces. The exposed, asymptomatic, symptomatic, and quarantine population is extinct when the fractional operator is 0.6 and below (see Figs. 3 – 6). Again, exposed, asymptomatic, symptomatic, and quarantine population is seen to reaches an early peak when the fractional operator value is reduced from 1. The number of exposed, asymptomatic, symptomatic, and quarantine individuals decays faster at the non-integer values. However, the number of immune individuals increases as $v$ reduces from 1.0 to 0.6 (see

Fig. 7). On the other hand, the recovered population is seen to have extinct at $v = 0.6$. This is so because the infections in the population have also extinct at the same value of the fractional operator (see Fig. 8).

## 6 Numerical Simulation of the Fractional Optimal Control

In this section, we analyze the numerical behavior of the fractional optimal control model using the parameter values given in Table 1, and the same initial conditions: $S(0) = 30800000, E(0) = 0, I_A(0) = 2, I_S(0) = 0, Q(0) = 0, V(0) = 0, R(0) = 0$. Using matlab OD45 Ruge-Kutta method, the results of the simulations are displayed in Figs. 9 – 20.

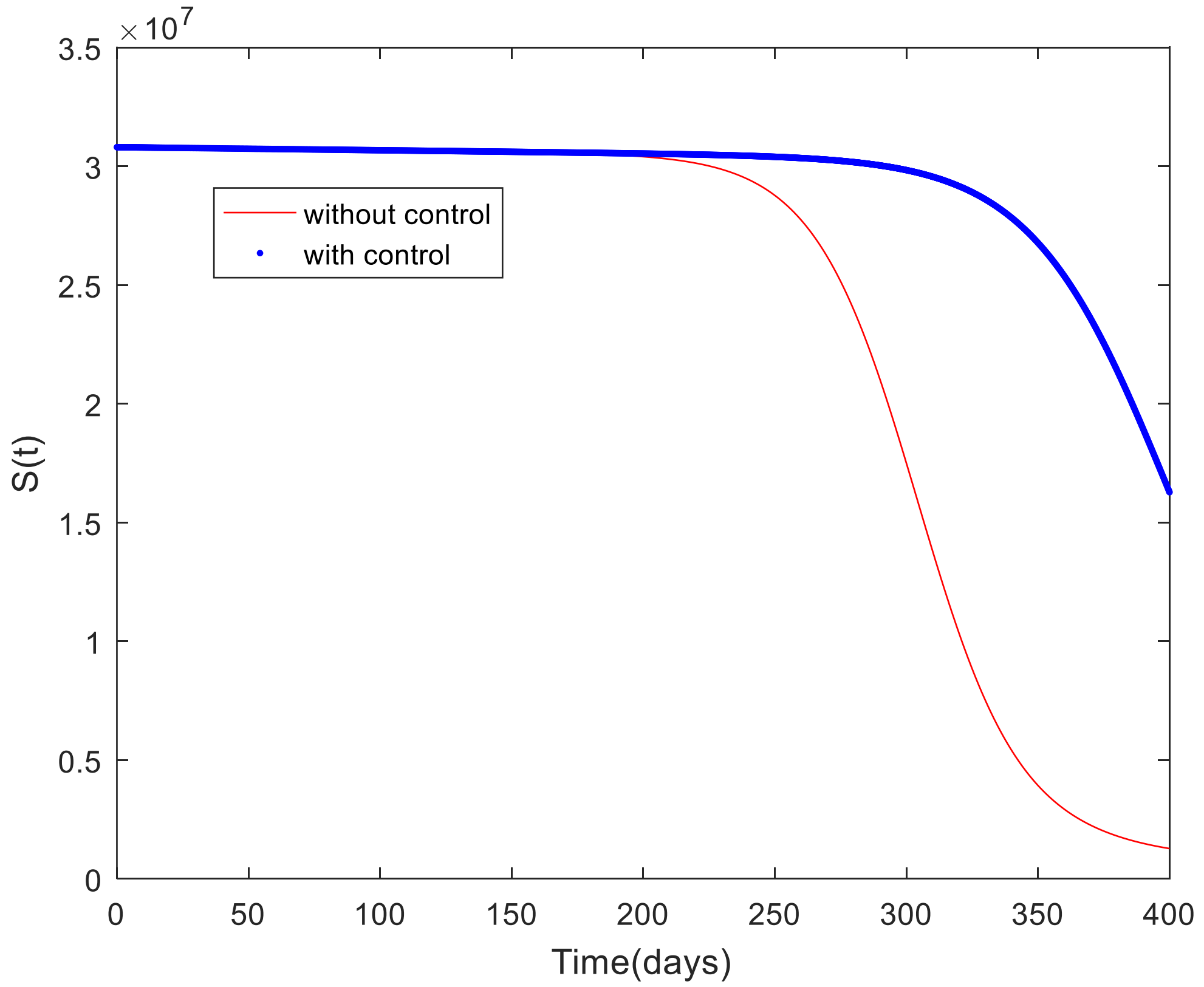

Fig. 9: Behaviour of the susceptible with and without control

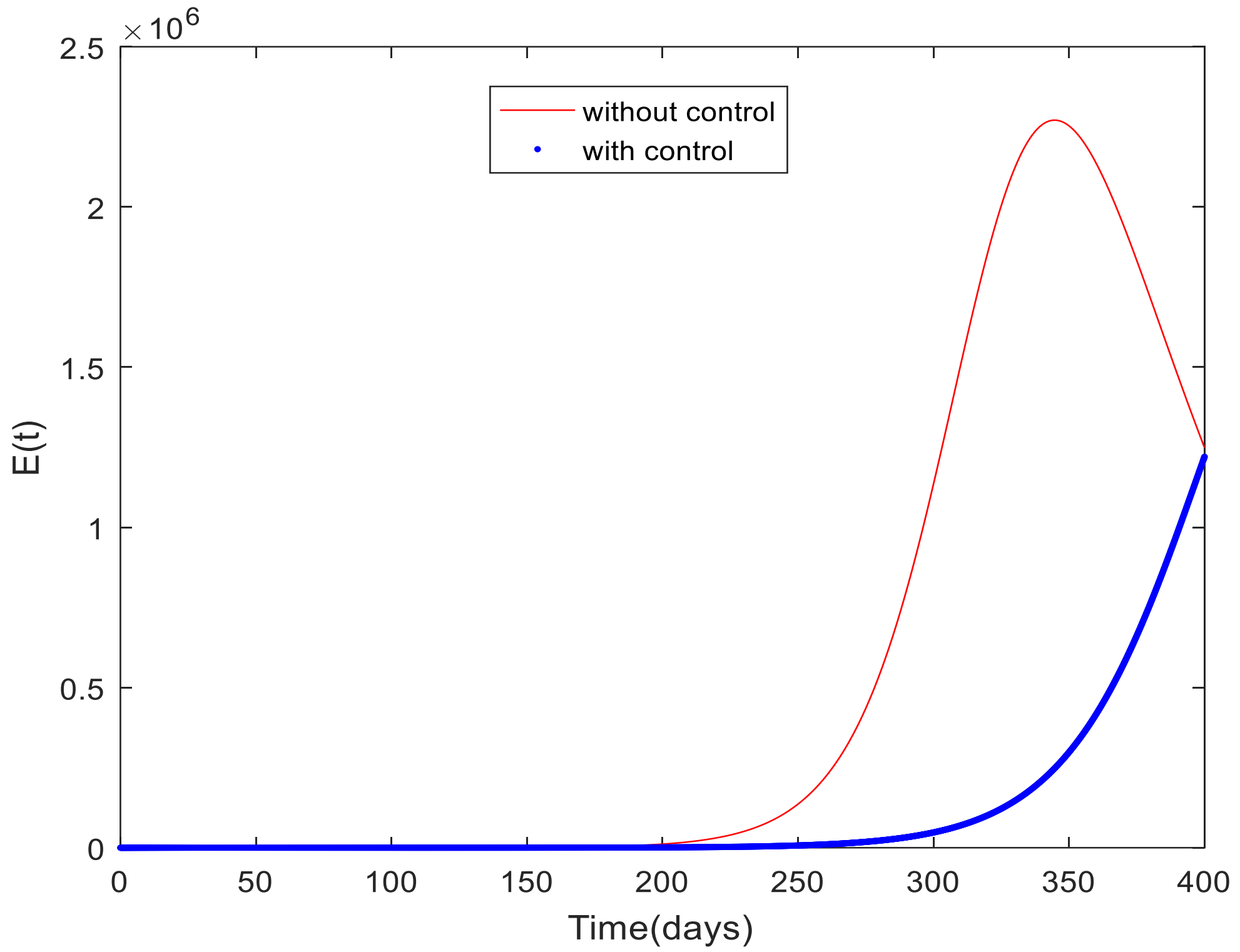

Fig. 10: Behaviour of the exposed individuals with and without optimal control

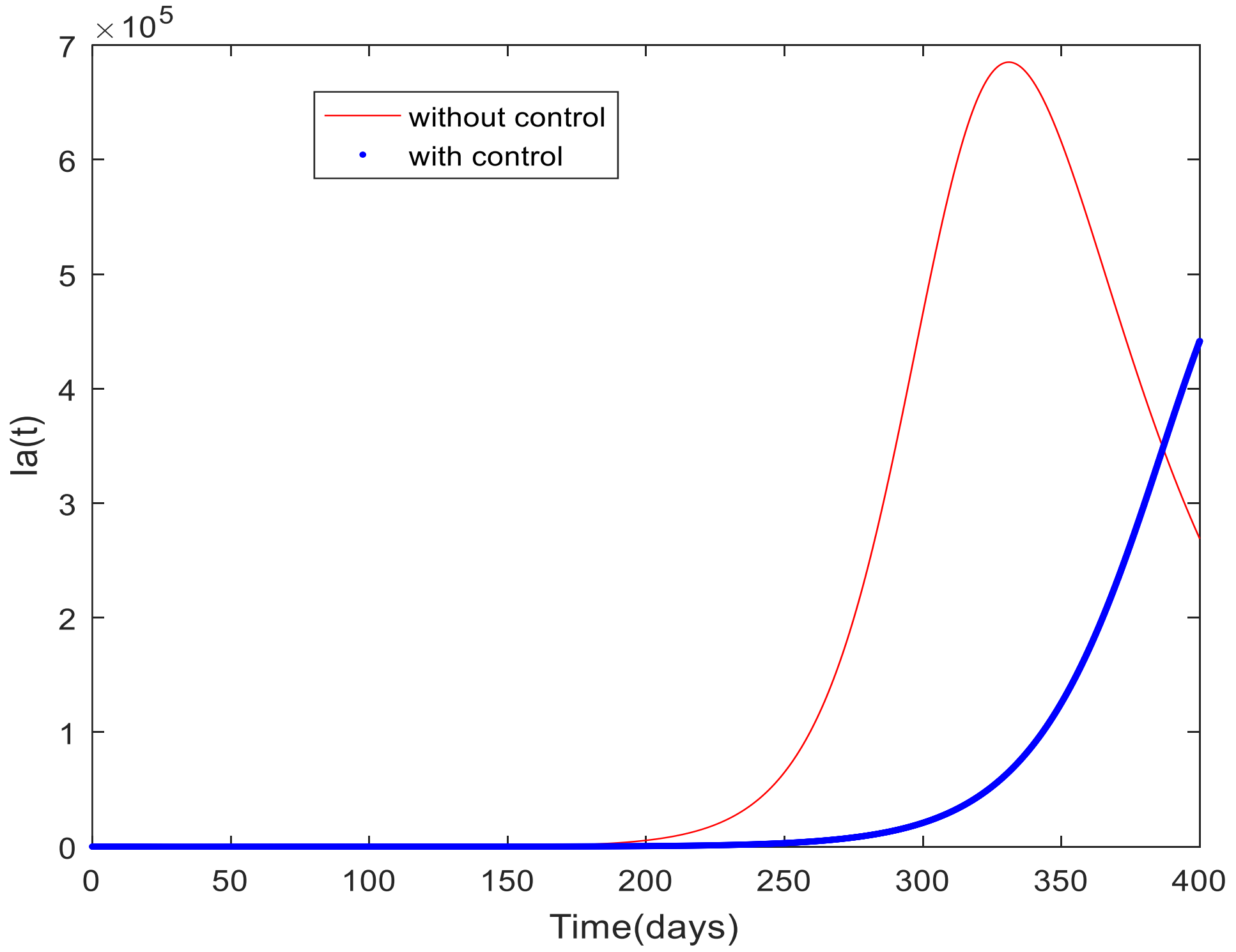

Fig. 11: Behaviour of the asymptomatic with and without control

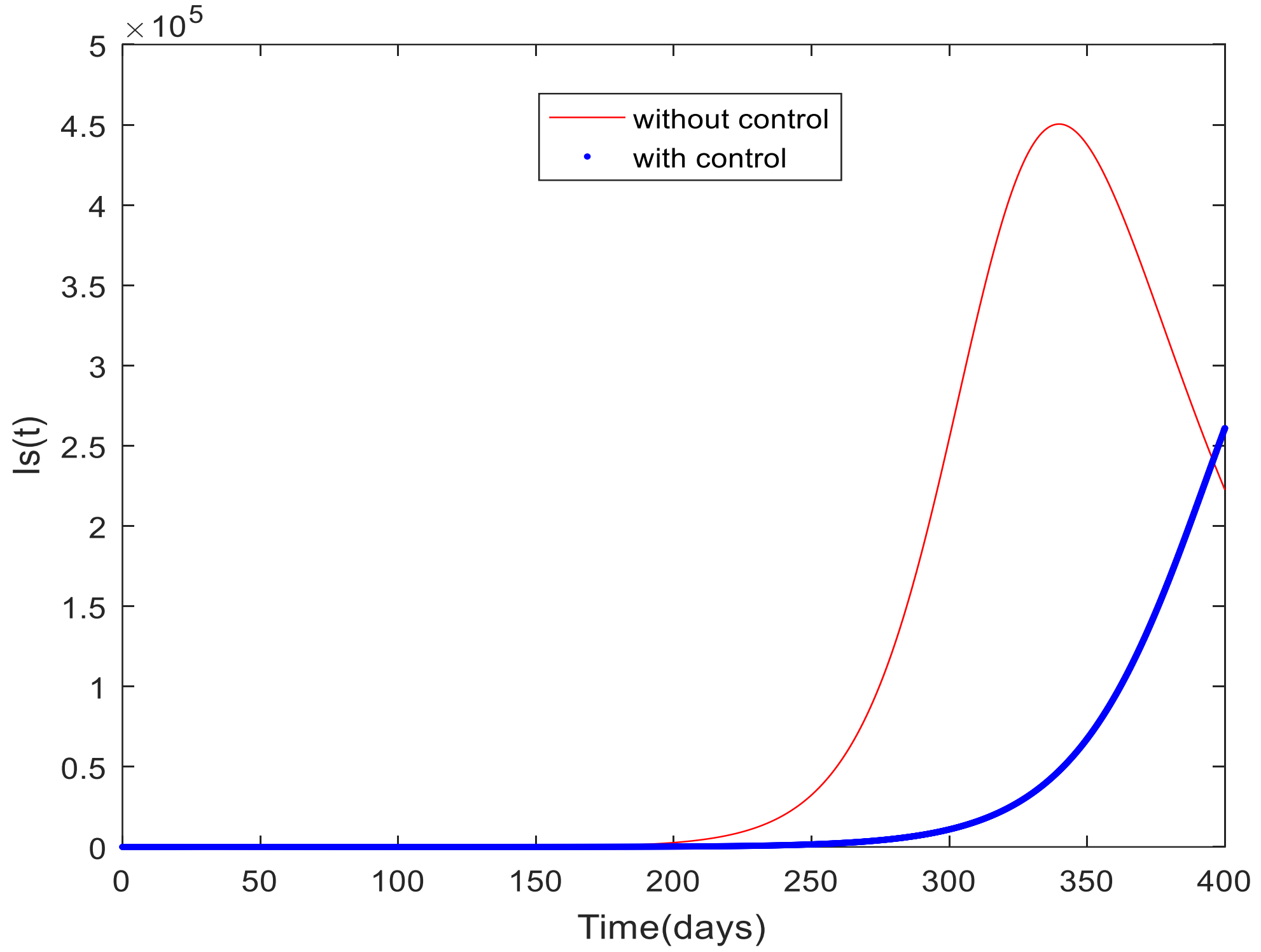

Fig. 12: Behaviour of the symptomatic with and without control

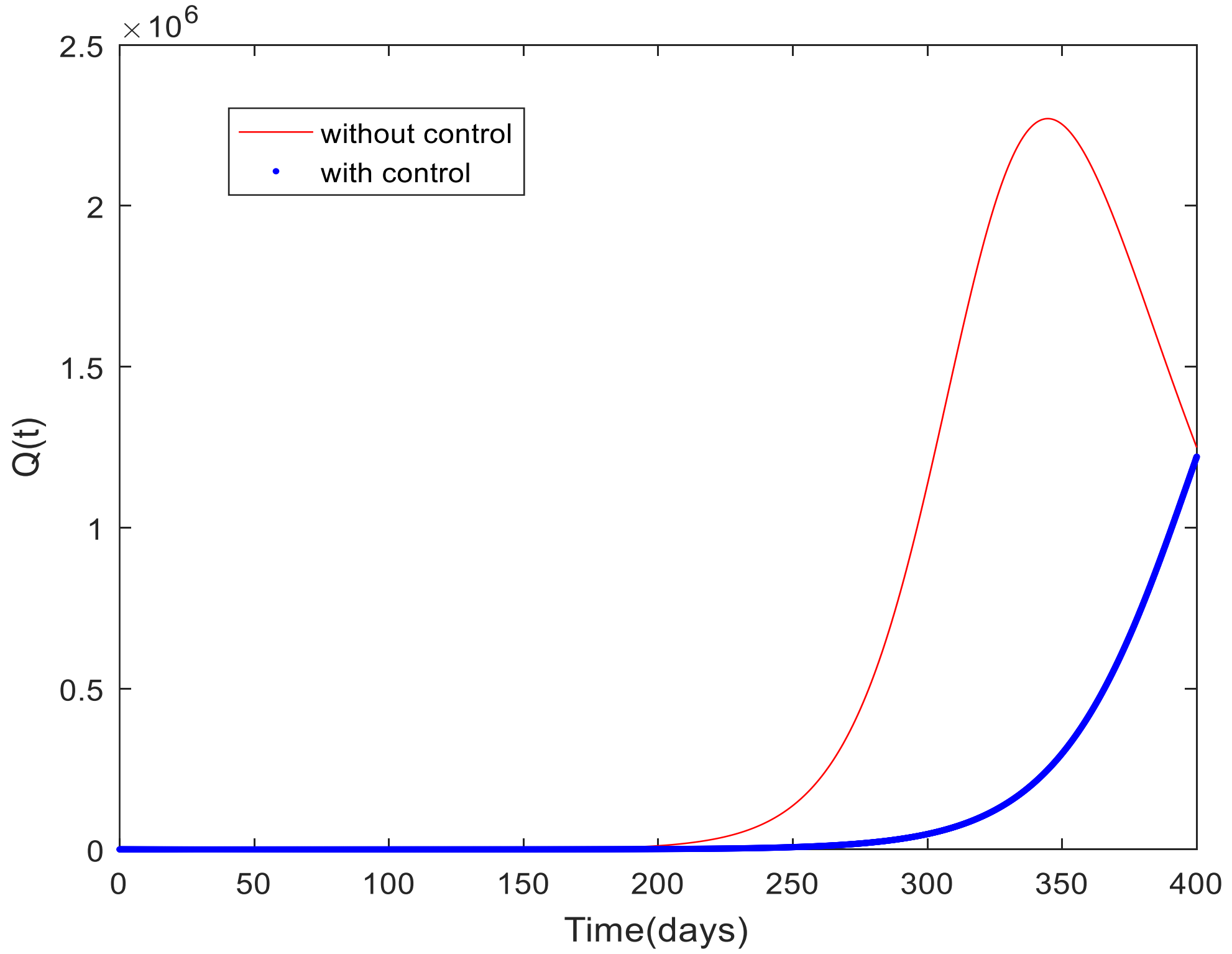

Fig. 13: Behaviour of the quarantine with and without control

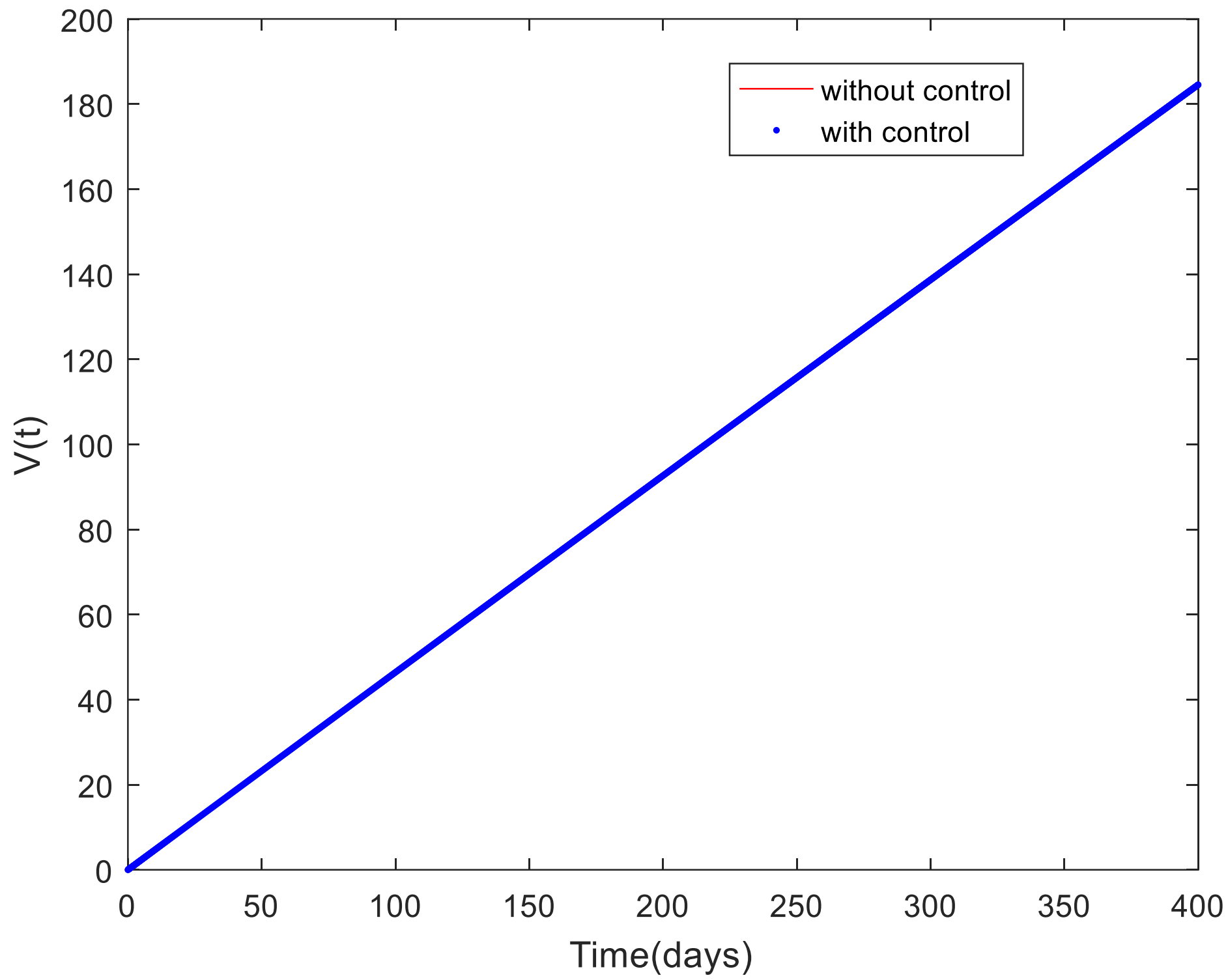

Fig. 14: Behaviour of the immune with and without control

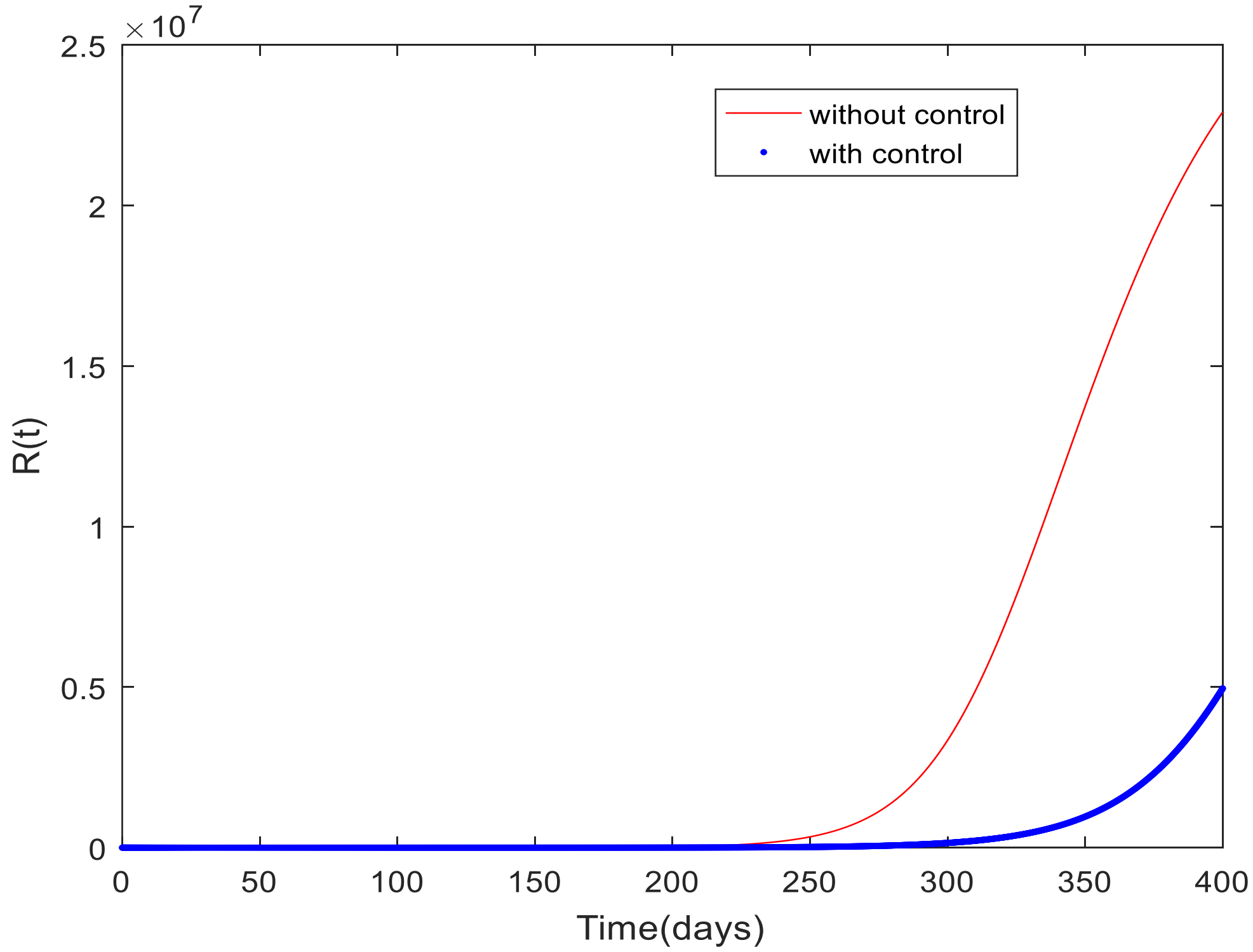

Fig. 15: Behaviour of the recovered with and without control

Figs. 9 – 15 depicts the behaviour of the susceptible, exposed, asymptomatic, symptomatic, quarantine, immune, and recovered individuals, respectively, when the optimal control $u_1$ is fully optimized whilst setting $u_2 = 0$ for the entire period of 400 days. With the social distancing control, it takes a longer period of 350 days before a significant decline in the susceptible population is observed as compared to a situation without optimal control which is 250 days (see Fig. 9). In Fig. 10, there is a drastic decline in the number of individuals that get exposed to the disease when social distancing is observed. This leads to a corresponding decline in the number of asymptomatic, symptomatic and quarantine individuals (see Figs. 11 – 13). The social distancing measures have no effect on the immune individuals as the population remains constant with or without optimal control (see Fig. 14). With a drastic fall in the number of infections in the case of an optimal control, the recovered population is also seen to decline when there is an optimal control (see Fig. 15). Figs. 16 – 22 describes the dynamics of each compartment when vaccination control is implemented.

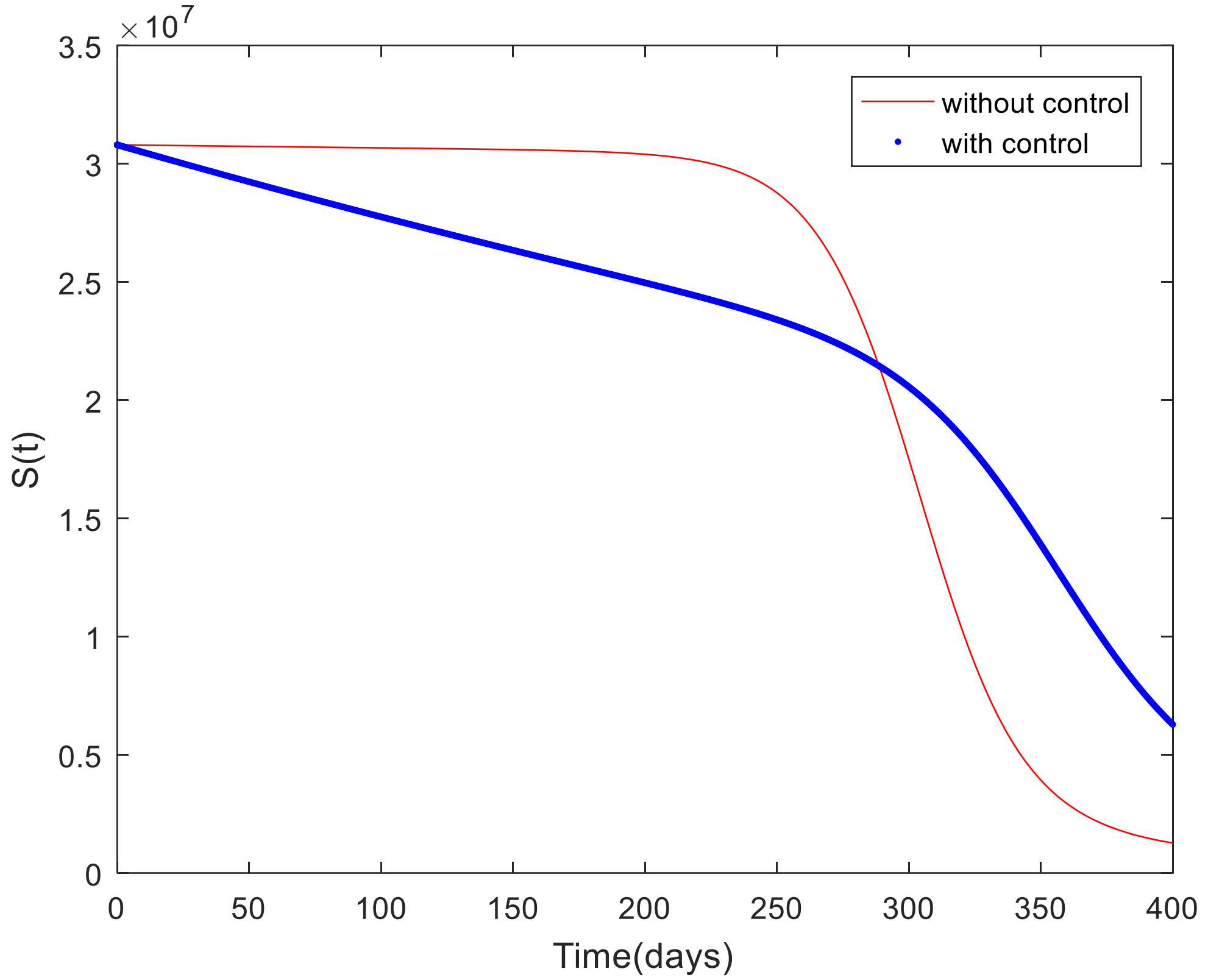

Fig. 16: Behaviour of the susceptible with and without vaccination control

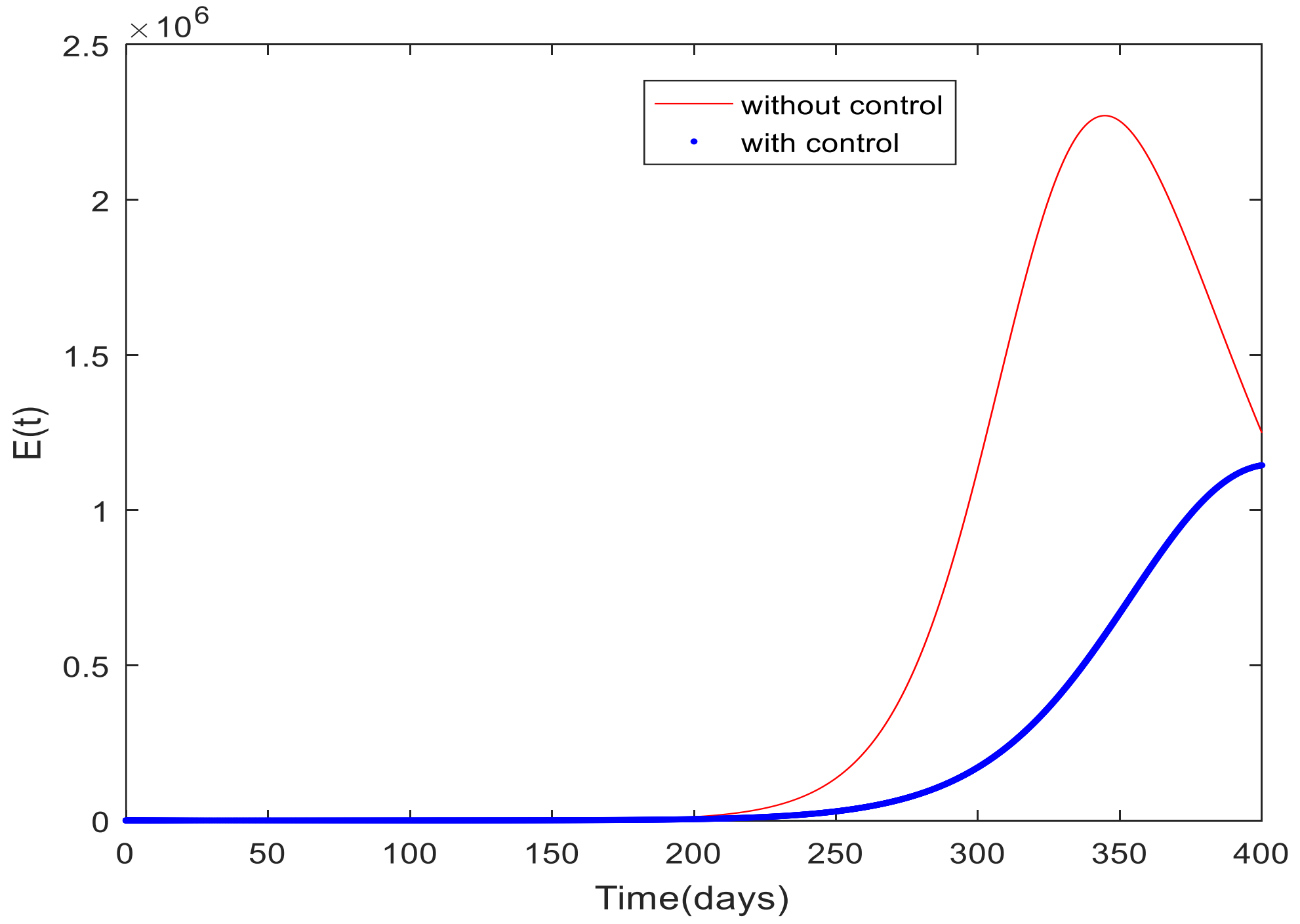

Fig. 17: Behaviour of the exposed individuals with and without vaccination control

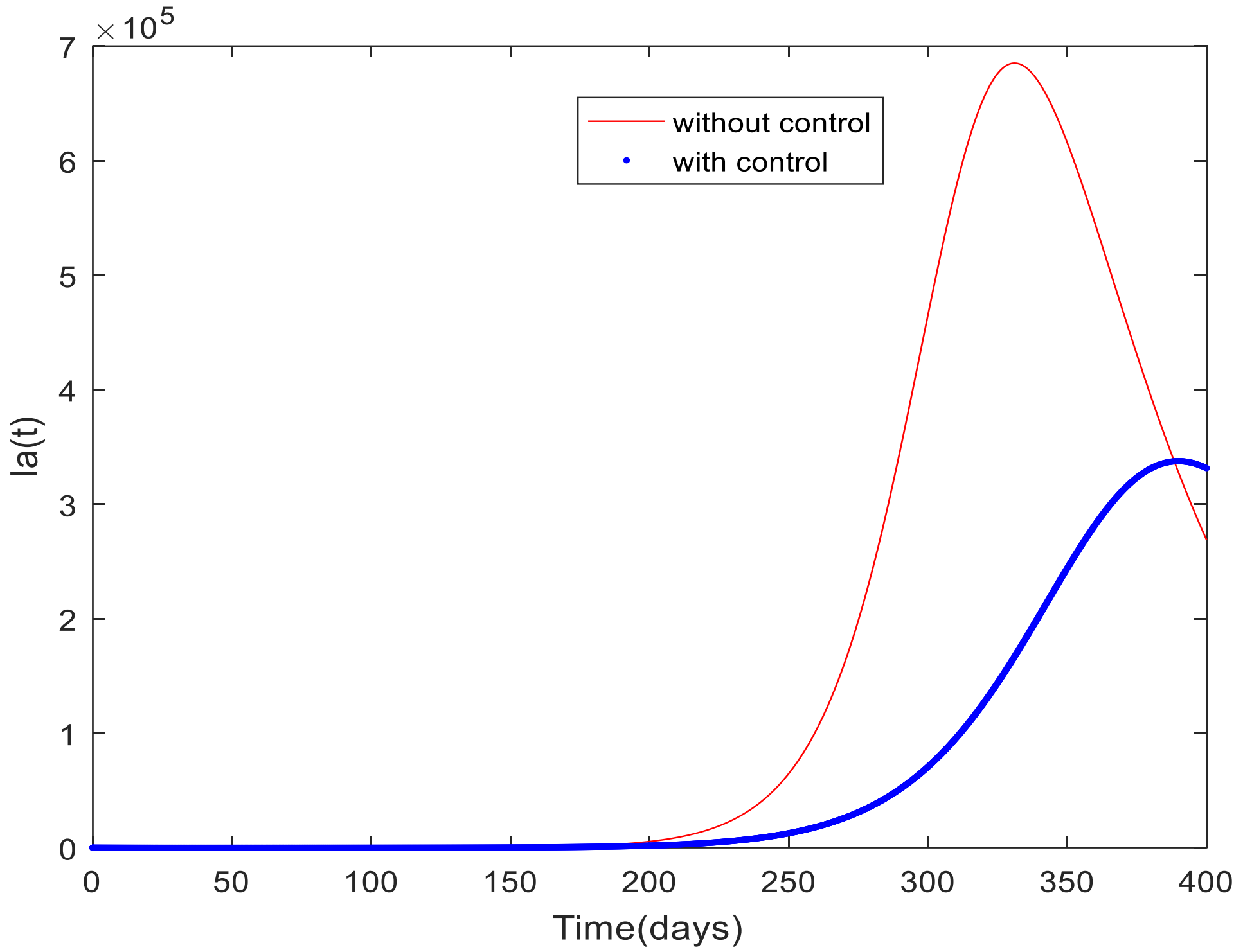

Fig. 18: Behaviour of the asymptomatic with and without vaccination control

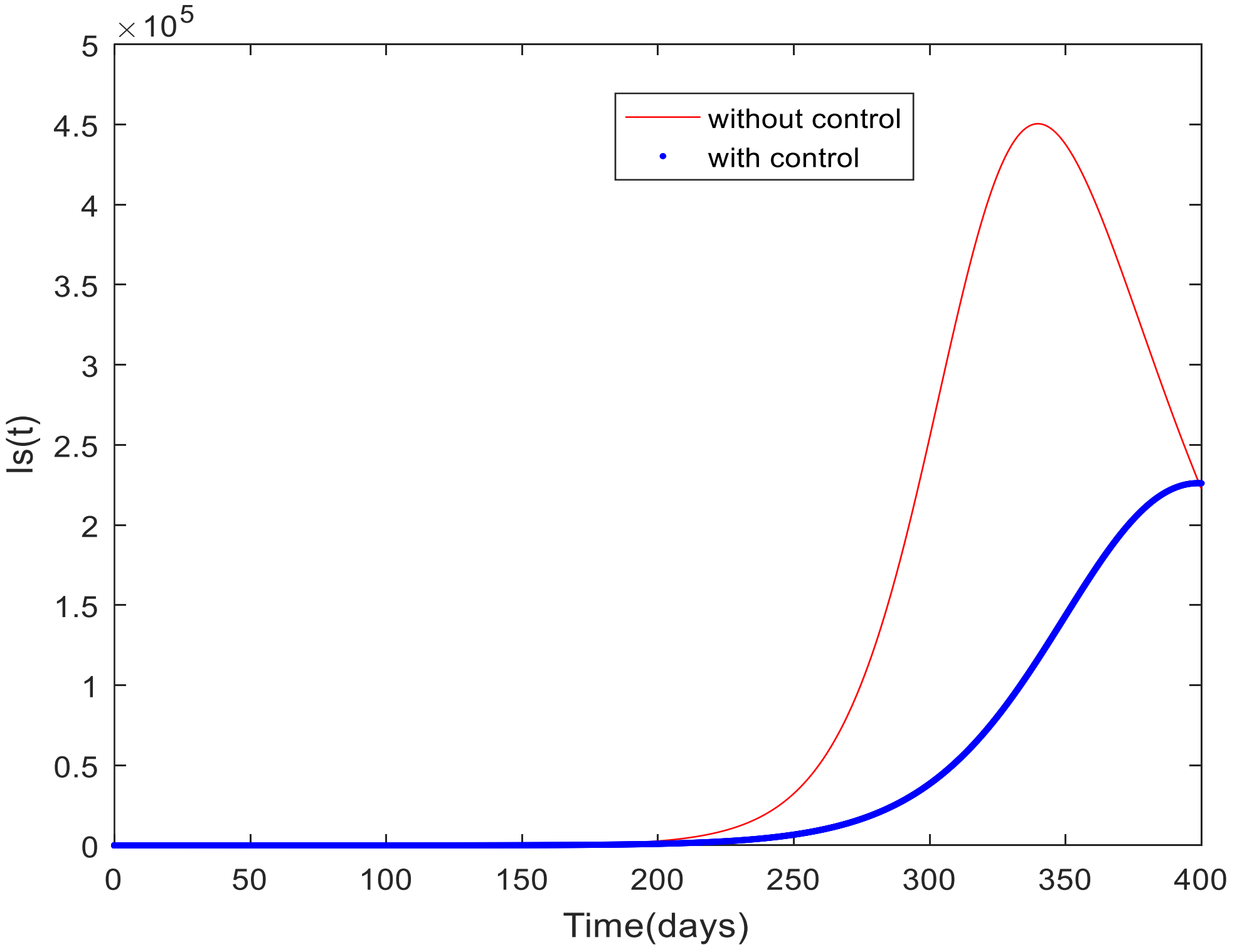

Fig. 19: Behaviour of the symptomatic with and without control

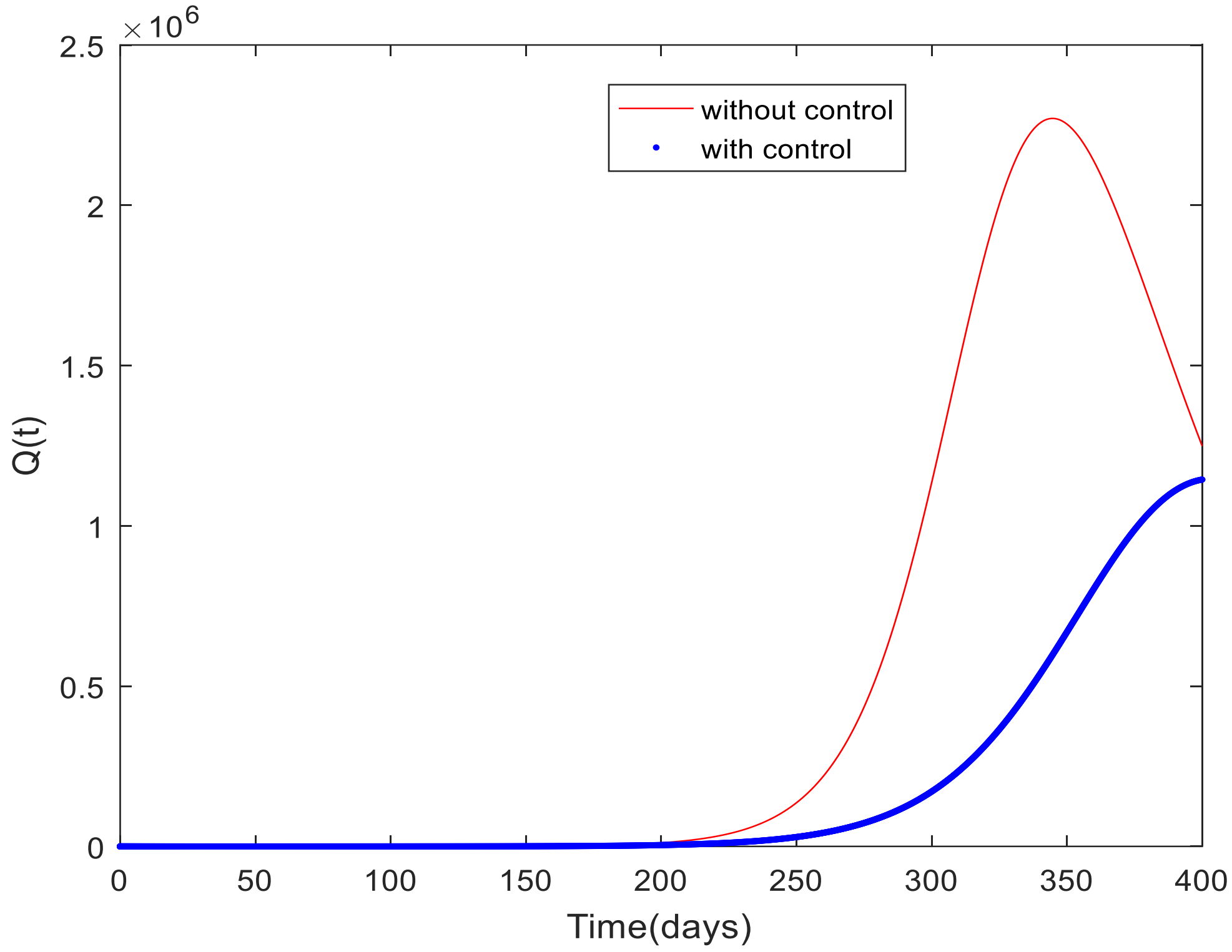

Fig. 20: Behaviour of the quarantine with and without vaccination control

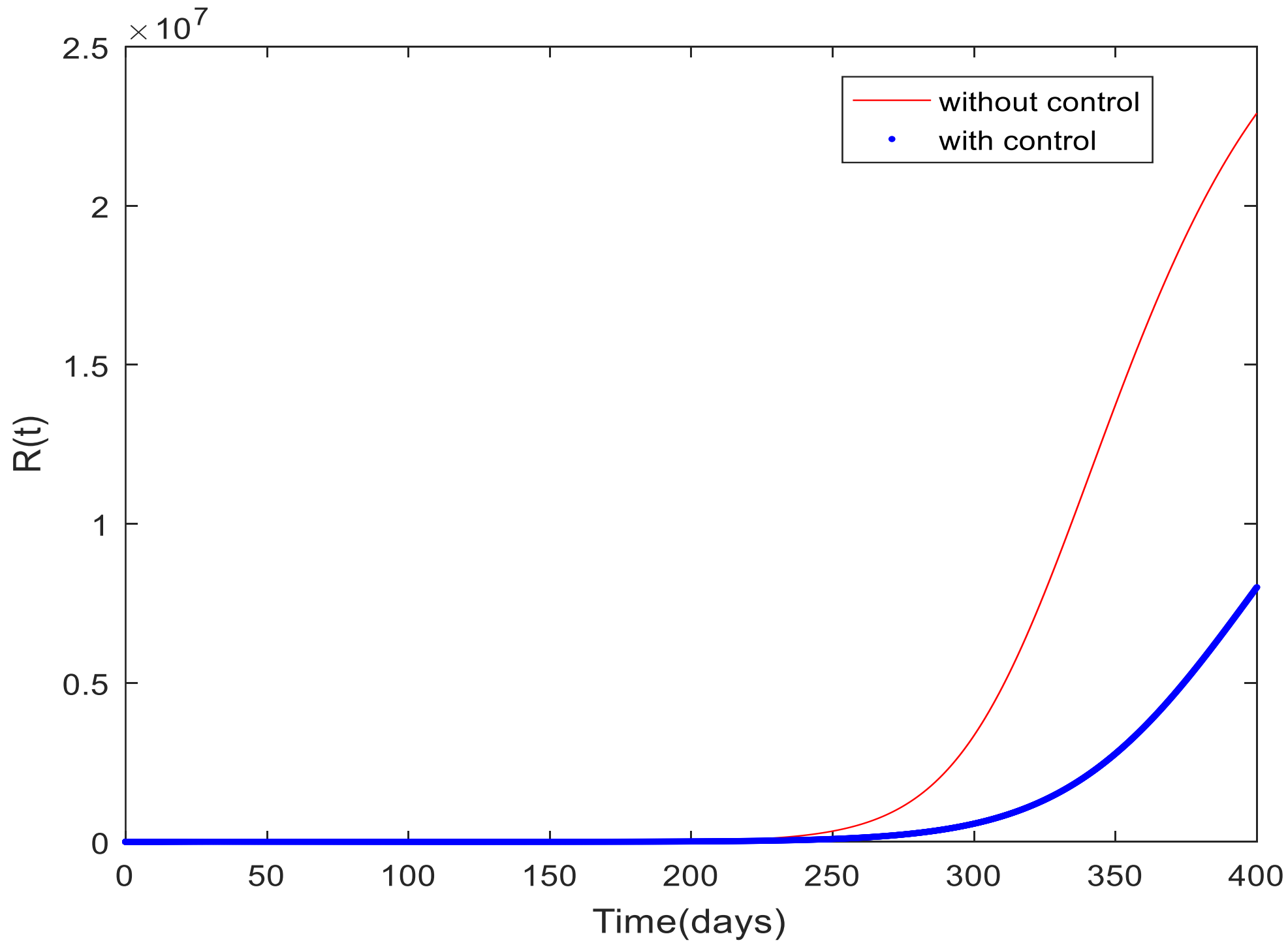

Fig. 21: Behaviour of the recovered with and without vaccination control

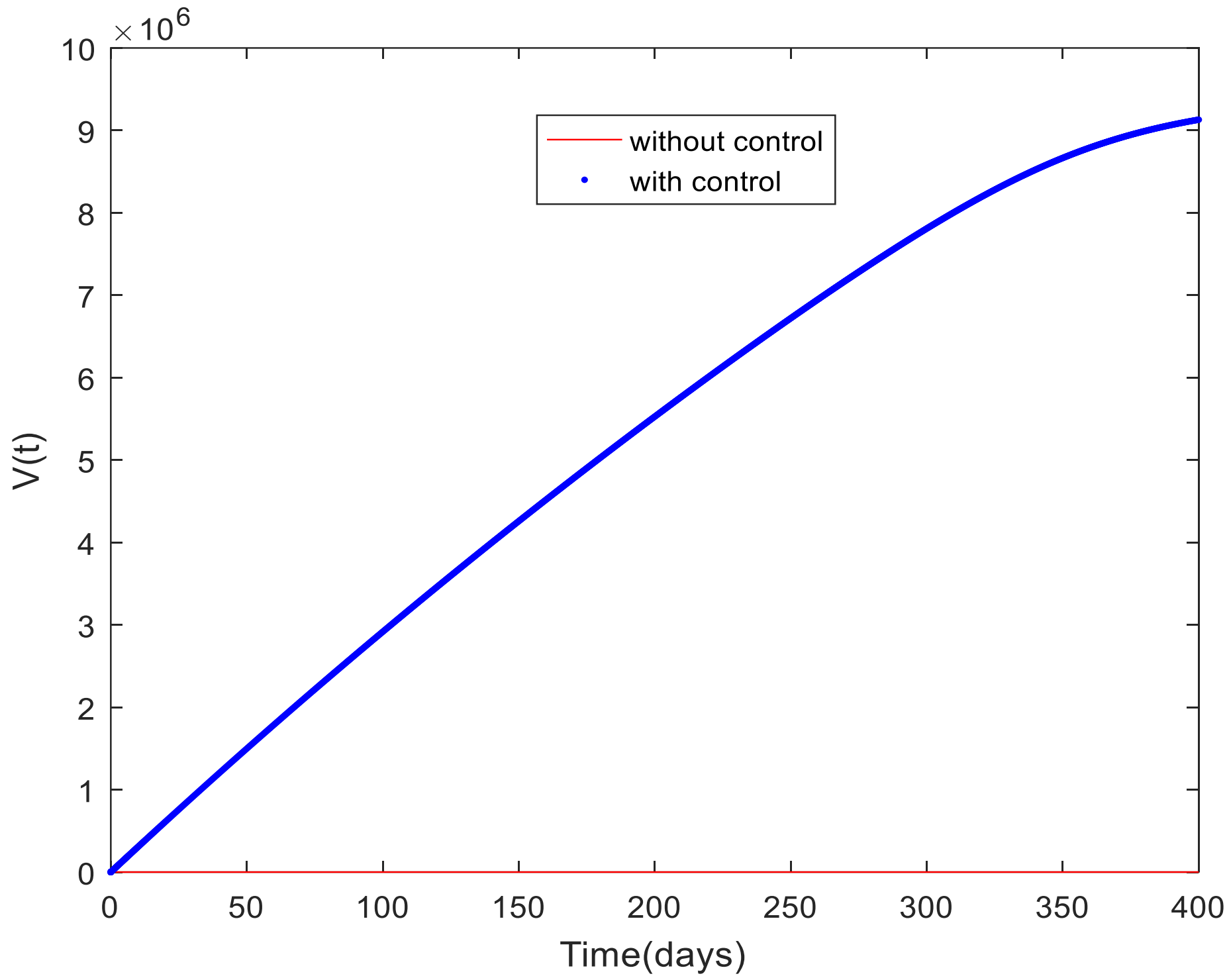

Fig. 22: Behaviour of the immune with and without vaccination control

Figs. 16 – 22 shows the behaviour of the susceptible, exposed, asymptomatic, symptomatic, quarantine, recovered and immune individuals respectively, when the vaccination control $u_2$ is fully optimized whilst setting $u_1 = 0$ for the entire period of 400 days. With the vaccination control, a significant decline in the number of susceptible individuals is observed for the first 300 days. However, after the 300 days there is a slow decline in the number when compared with a situation without optimal control. The vaccination reduces the number of individuals susceptible to the disease for the first 300 day (see Fig. 16). Vaccinating the susceptible individuals leads to a decline in the number of exposed, asymptomatic, symptomatic, and quarantine individuals as compared with a situation without the optimal control (see Fig. 17 – 20). The decline in infections leads to a decline in the recovered population (see Fig. 21). There is a high number of individuals that develop immunity to the disease due to the vaccination of susceptible individuals (see Fig. 22).

## 7 Conclusion

In this study, the model in [20] has been modified and formulated using the fractional-order derivative defined in Atangana – Baleanu – Caputo sense. The basic properties such as the equilibrium points, basic reproduction number, and uniqueness of the solutions have been explored. Fractional optimal controls were incorporated into the model to determine appropriate intervention strategies in curbing the spread of the disease. The model was validated using the parameter values given in [20] for a period of 400 days. Matlab software fourth-order Runge-Kutta method was used for the simulations. Results of the numerical simulation shows there is a significant number of individuals who become exposed, asymptomatic, symptomatic, quarantine, and recovered when the fractional operator $v$ is above 0.6. The number of immune individuals increases with a reduction in the fractional operator value from 1 to 0.6. Contrary to the number of immune individuals, the number of susceptible individuals declines as the fractional operator value decreases. The numerical simulation of the optimal control model demonstrates that vaccination and social isolation are both very successful strategies for preventing the spread of the disease. Social isolation had no impact on the immune population, while vaccination controls produced a sizable proportion of disease-immune individuals.

**Data Availability**

The data/information supporting the formulation of the mathematical model in this paper are/is from the Ghana Health Service website: https://www.ghs.gov.gh/covid19/ which has been referenced in the manuscript [39].

**Declaration of interest**

No conflict of interest regarding the content of this article

**Funding**

The research did not receive funding from any source.

**Acknowledgement**

This work was submitted as part of a student thesis at Kwame Nkrumah University of Science and Technology, Ghana. Previous versions of the work are also available as a pre-print on https://arxiv.org/ftp/arxiv/papers/2201/2201.08689.pdf [40] and https://arxiv.org/ftp/arxiv/papers/2201/2201.11659.pdf [15] .